\documentclass[twocolumn]{aastex7}

\usepackage{graphicx}	
\usepackage{xcolor}
\usepackage{anyfontsize}
\usepackage{xspace}
\usepackage{newtxmath}
\usepackage{newtxtext,newtxmath}
\usepackage{subcaption}

\newcommand\cxo{\textit{Chandra}\xspace}
\newcommand\xmm{\textit{XMM-Newton}\xspace}

\shorttitle{Disentangling the Magnetospheric and Colliding-Wind X-ray/Radio Emission from Plaskett's Star}
\shortauthors{Biswas et al.}

\begin{document}

\title{Disentangling Magnetospheric and Colliding-Wind X-ray and Radio Emission in the Post-Mass-Transfer Magnetic Massive Binary Plaskett's Star}

\author[0000-0002-1741-6286]{Ayan Biswas}
\affiliation{National Centre for Radio Astrophysics, Tata Institute of Fundamental Research, Ganeshkhind, Pune-411007, India }
\affiliation{Department of Physics, Engineering Physics \& Astronomy, Queen’s University, Kingston, Ontario K7L 3N6, Canada}
\affiliation{Department of Physics \& Space Science, Royal Military College of Canada, PO Box 17000, Station Forces, Kingston, ON K7K 7B4, Canada}
\email[show]{abiswas@ncra.tifr.res.in, ayan.scc2@gmail.com}

\author[0000-0002-1854-0131]{Gregg A. Wade}
\affiliation{Department of Physics \& Space Science, Royal Military College of Canada, PO Box 17000, Station Forces, Kingston, ON K7K 7B4, Canada}
\affiliation{Department of Physics, Engineering Physics \& Astronomy, Queen’s University, Kingston, Ontario K7L 3N6, Canada}
\email{Gregg.Wade@rmc.ca} 

\author[0000-0002-3331-7595]{Maurice Leutenegger}
\affiliation{NASA/Goddard Space Flight Center, Code 662, Greenbelt, MD 20771, USA}
\email{maurice.a.leutenegger@nasa.gov}  

\author[0000-0002-0844-6563]{Poonam Chandra}
\affiliation{National Radio Astronomy Observatory, 520 Edgemont Road, Charlottesville VA 22903, USA }
\email{pchandra@nrao.edu}  

\author[0000-0003-4071-9346]{Yaël Nazé}\thanks{F.R.S.-FNRS Senior Research Associate}
\affiliation{Groupe d’Astrophysique des Hautes Energies, STAR, Université de Liège, B5c, Allée du 6 Août 19c, B-4000 Sart Tilman, Liège, Belgium}
\email{ynaze@uliege.be}  

\author[0000-0001-8704-1822]{Barnali Das}
\affiliation{National Centre for Radio Astrophysics, Tata Institute of Fundamental Research, Ganeshkhind, Pune-411007, India }
\email{barnali@ncra.tifr.res.in} 

\author[0000-0002-5633-7548]{Veronique Petit}
\affiliation{Department of Physics and Astronomy, Bartol Research Institute, University of Delaware, Newark, DE 19716, USA}
\email{VPetit@udel.edu}


\correspondingauthor{Ayan Biswas}


\begin{abstract}

Plaskett's star is an O+O binary system harboring a kG-order magnetic field associated with the broad-lined secondary. It is the only O-type star believed to host a centrifugally supported magnetosphere. We present orbital- and rotational-phase resolved X-ray and low-frequency radio observations of Plaskett's star. X-ray observations were taken with \cxo\ using the High Energy Transmission Grating Spectrometer (HETGS), while the radio observations come from uGMRT and VLA. The X-ray light curve shows rotational modulation according to the magnetic component's rotation period, with maximum flux near magnetic minima. We confirm the decline in flux near orbital phase 0.5 where the slow rotator primary is in front of its magnetic companion. We used the $f/i$ ratios of \ion{Si}{13} and \ion{Mg}{11} from high-resolution X-ray spectroscopy to infer the location of the plasma emitting X-rays. We find no statistically significant variation in line widths or centroid shifts with either orbital or rotational phase; however, given instrumental limitations, this non-detection may partly reflect observational constraints, while still suggesting a complex and extended X-ray emitting plasma distribution. We could not detect the star in the low-frequency radio bands (0.6--1.5 GHz). Re-analysis of existing Very Large Array data reveals an additional detection at 3 GHz, consistent with a nonthermal contribution predicted by magnetospheric models, although no direct signature was obtained. At low frequencies, nonthermal emission from the magnetosphere and the colliding-wind region is likely suppressed by free-free absorption in the winds of one or both components.

\end{abstract}



\section{Introduction} \label{sec:intro}

O-type stars are the hottest {main sequence} stars with high luminosities that light up and ionize the nearby interstellar medium and drive strong, high-speed stellar winds, making them the nucleosynthetic engines of galaxies \citep{Crowther2010}. Approximately 10\% O stars have been found to harbor strong, stable, and primarily dipolar magnetic fields (strength $\sim$kG, \citealt{Grunhut2017, Petit2019}). The detected magnetic fields are characteristically different from those of cool, low-mass stars \citep{Donati2009}. The widely favored hypothesis is that these fossil fields are either remnants of the galactic field accumulated and possibly enhanced by a dynamo field generated in an earlier phase of evolution or generated through the merger process \citep{Ferrario2008, Schneider2019}. These strong organized magnetic fields can confine the radiatively driven stellar winds (e.g. \citealt{Babel1997}), thus quenching the mass loss of the star \citep{ud-Doula2009}, and in turn affecting the final mass of the star towards the end of main sequence. The quenching of mass-loss can ultimately affect the fate of these stars after their death and explain the origin of unusually large solar mass black holes \citep{Petit2013}. The confined stellar winds in these stars create a magnetosphere that can radiate in various wavebands \citep{Andre1988, Linsky1992, Petit2013, Naze2014}.

Among O-type stars with detected magnetic fields, the fast rotator in the close binary Plaskett's star has a variety of unusual properties. Plaskett's star (HD 47129) is a bright (V = 6.06) spatially unresolved system with two distinct chemically peculiar O-type components. The slowly rotating primary is an O8III/I type star with narrow spectral {lines}, and the fast rotating secondary is an O7.5V/III type star showing broad spectral lines \citep{Linder2008}. Plaskett's star's broad-line component is a fairly typical magnetic early-type star, with a stable surface magnetic field ($B_d = 850$ G) that is roughly a tilted dipole, but with discernible deviations from a pure dipole \citep{Grunhut2013, Grunhut2022}. The Alfv{\'{e}}n radius ($R_{\rm A}$) of the magnetic component is greater than the Kepler co-rotation radius ($R_{\rm K}$). Thus, the star's wind is directed and confined by the magnetic field, creating a structured centrifugal magnetosphere (CM) that is apparent through optical emission lines \citep{Petit2013}. Although a CM is very common among B-type stars, Plaskett's star is the only known magnetic O-type star to possess a CM \citep{Petit2013}.

\cite{Grunhut2022} attempted to map the magnetic field of the star using Zeeman-Doppler Imaging (ZDI) which led them to consider models in which the radial velocity (RV) of the broad-line star was constant. They were unable to fit the Stokes V profiles while simultaneously considering the RV variation of the broad-line star according to the 14.4 day orbital period of the system, as inferred by \cite{Linder2008}. { \cite{Wade2026} have recently revisited the RV variability of the rapidly rotating magnetic component. By isolating the O\,{\sc iii} $\lambda5592$ line, which is formed predominantly in the atmosphere of the magnetic secondary, they detected a weak RV variation with a semi-amplitude of $K_2 = 29.6 \pm 4.5$ km\,s$^{-1}$ at the orbital period. Their orbital solution gives a mass ratio $q=M_2/M_1=6.9^{+1.2}_{-0.9}$, with minimum masses of $2.4\pm0.5$ and $16.8\pm0.7\,M_{\odot}$ for the primary and secondary, respectively. Assuming spin--orbit alignment, they infer absolute masses of $M_1=5.9^{+1.8}_{-1.4}\,M_{\odot}$ and $M_2=40.8^{+9.3}_{-6.7}\,M_{\odot}$, and an orbital separation of $a=78.4^{+5.5}_{-4.5}\,R_{\odot}$. These results strongly support a post-mass-transfer configuration in which the narrow-lined primary is the stripped donor and the rapidly rotating magnetic secondary is the mass-gaining star \citep{Wade2026}.}

{Some of the relevant stellar and orbital parameters of Plaskett's star are given in Table \ref{tab:parameters}. The orbital inclination adopted by \cite{Wade2026}, $i_{\rm orb}=48\pm4^\circ$, is based on the assumption that the rotation and orbital axes of the secondary are aligned, rather than being a direct measurement of the orbital inclination. The magnetic geometry of the secondary is particularly important for interpreting the X-ray and radio emission. The secondary has a large-scale approximately dipolar magnetic field with a polar strength of and a highly oblique magnetic axis, while its Alfv\'en radius extends several stellar radii from the stellar surface \citep{Grunhut2013, Grunhut2022}. The combination of rapid rotation and strong magnetic confinement therefore produces a centrifugal magnetosphere that extends into the region where the primary wind is expected to interact with the secondary. This makes Plaskett's star distinct from a conventional colliding-wind binary and motivates considering both magnetically confined wind shocks and primary-wind--magnetosphere interaction when interpreting its X-ray emission.}

\begin{deluxetable}{lll}
\tablecaption{Different stellar and orbital parameters of Plaskett's star. { $T_{0, \rm orb}$ is defined as the reference time when the fast rotating secondary is in front of the slowly rotating star. $T_{0, \rm rot}$ is the reference time for photometric maxima when the north magnetic hemisphere of the star is visible. $a \sin{i}$  is the projected semi-major axis, $R_{\rm RL}$ is the Roche-lobe radius, and $M_{\rm spec}$ is the spectroscopic mass. {The rotational inclination is the angle between the secondary's rotation axis and the line-of-sight.} } \label{tab:parameters}}
\tablehead{
\colhead{Parameters (Units)} & \multicolumn{2}{c}{Value} \\
\colhead{} & \colhead{}{Primary}  & \colhead{Secondary}
}
\startdata
 \multicolumn{3}{c}{\textit{Spectral Parameters \citep{Linder2008}}} \\
Spectral Type &  O8 III/I &  O7.5 V/III \\
$T_{\rm eff}$  (K)  & $33500 \pm 2000$ & $33000 \pm 2000$ \\
$\log {L}$  ($L_{\odot}$)  & $5.35$ & $5.09$ \\
$\log {g}$ & $3.5 \pm 0.1$ & $3.5 \pm 0.1$ \\
$v_{\infty}$  (km/s) & 2450 & 3500 \\
\multicolumn{3}{c}{\textit{Magnetospheric Parameters \citep{Grunhut2022}}} \\
Rotation Period ($P_{\rm rot}/{\rm [d]}$)  &  ... & $1.21551^{+0.00028}_{-0.00034}$  \\
$T_{0, \rm rot}$ (MJD) & ... & $55960.500^{+0.048}_{-0.034}$ \\
{ Rot. (/Orb.)} Inclination, i ($^{\circ}$) & ...  & $48 \pm 4$ \\
Obliquity, $\beta$ ($^{\circ}$) & ... & $\sim 89$ \\
$B_{\rm P}$ (kG) & ... & $\sim 0.85$ \\
$R_{\rm A}$ ($R_{*})$  & ... & $\sim 4.4$ \\
$R_{\rm K}$ ($R_{\rm eq})$  & ... & $1.55 \pm 0.2$ \\
 \multicolumn{3}{c}{\textit{Orbital and Stellar Parameters from \cite{Wade2026}}} \\
 Orbital Period ($P_{\rm orb}/{\rm [d]}$)  &  \multicolumn{2}{c}{$14.39626 \pm  0.000953$}  \\
$T_{0, \rm orb}$ (MJD) & \multicolumn{2}{c}{$55966.19 \pm 0.06$} \\
Assumed eccentricity (e) &   \multicolumn{2}{c}{0.0} \\
Distance (pc) & $1270 \pm 96$ \\
$M \sin^3{i}$ ($ M_{\odot}$) & $2.4 \pm 0.5$  & $16.8 \pm 0.7$ \\
$a \sin{i}$ ($R_{\odot}$) &  $58.4 \pm 0.5$ &  $8.4 \pm 1.3$ \\
$R_{\rm RL}$ ($R_{\odot}$) &  $20.6_{-1.7}^{+1.9}$ &  $49.3_{-2.8}^{+3.5}$ \\
$R_{*}$ ($R_{\odot}$) &  $12.3 \pm 1.2$ &  $10.4 \pm 1.1$ \\
$M_{\rm spec}$ ($M_{\odot}$) &  $5.9_{-1.4}^{+1.8}$ &  $40.8_{-6.7}^{+3.5}$ \\
$T_{\rm eff}$  (kK)  & $31 \pm 1$ & $34 \pm 2$ \\
$\log {L}$  ($L_{\odot}$)  & $5.10 \pm 0.08$ & $5.12 \pm 0.11$ \\
$\log {g}$ & $3.3 \pm 0.2$ & $4.0 \pm 0.3$ \\
\enddata
\end{deluxetable}

Plaskett's star shows significant photometric variability \citep{Mahy2011, Burssens2020}. From \textit{CoRoT} data, \cite{Mahy2011} found the most significant frequencies to be $f1 =$  1.646 d$^{-1}$ (or $p1 \sim 0.6075$ days) and $f2= 2 \times f1 =$  0.823 d$^{-1}$. \cite{Burssens2020} reported a similar frequency in the \textit{TESS} data set ($\nu$ = 0.811(3) d$^{-1}$), which is close to the \textit{CoRoT} frequency. Later, \cite{Grunhut2022} discovered that the origin of these frequencies is the rotationally modulated, magnetically confined wind plasma of the star. Both \textit{CoRoT} and \textit{TESS} light curves show a double-wave variation with significant cycle-to-cycle variability \citep{Grunhut2022, Stacey2022}. \cite{Grunhut2022} noticed that the \textit{Hipparcos} light curve of this target does not phase well with the \textit{TESS} or the \textit{CoRoT} data. However, the sparse time sampling, high uncertainties on each point of \textit{Hipparcos} data make it unreliable for period analysis.

\begin{figure}
    \centering
    \includegraphics[width=\linewidth]{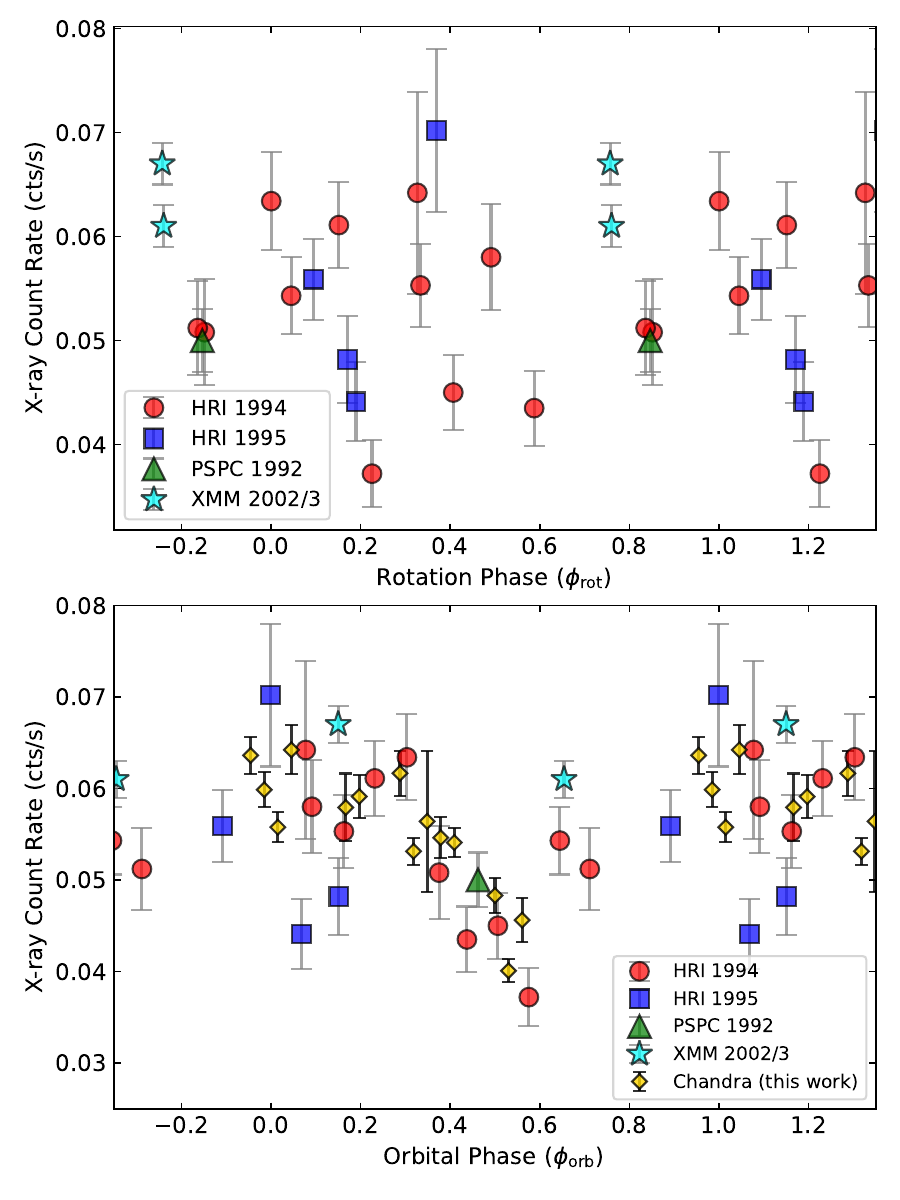}
    \caption{Variation of X-ray count rate from the Plaskett's star on rotational (top)  and orbital (bottom) timescales, with observations found in the literature, and scaled to ROSAT count rates. The data is taken from \cite{Linder2006} and this work, and phased according to the rotational ephemeris obtained by \cite{Grunhut2022}, and orbital ephemeris obtained by \cite{Wade2026}. { $\phi_{\rm orb}=0$ is defined as the reference phase when the fast rotating secondary is in front of the slowly rotating star. $\phi_{\rm rot}=0$ is the reference time for photometric maxima when the north magnetic hemisphere of the star is visible.}
    }
    \label{fig:linder}
\end{figure}

Plaskett's star was first detected in X-rays by the \textit{ROSAT} All-Sky Survey \citep{Berghoefer1996}. \cite{Linder2006} followed up the target at two epochs with the \textit{XMM–Newton}, with a total observing time of $\sim 45$ ks. Their spectral analysis of the XMM-{EPIC} spectra suggests that the X-ray spectrum is best fitted by a two-temperature thermal plasma model along with a power-law component ($kT_1 \simeq 0.64$, $kT_2 \simeq 1.25$, and $\Gamma \simeq 2.6$). The authors did not find short-term variability on timescales of a few minutes to about an hour. They also reported a possible roughly sinusoidal modulation of the X-ray count rate based on their reanalysis of \textit{ROSAT} data, which peaked around conjunction with the fast rotator in front of the slow-rotating primary. We plotted all the X-ray data reported by \cite{Linder2006}, and phased according to the latest rotational ephemeris obtained by \cite{Grunhut2022}, and orbital ephemeris obtained by \cite{Wade2026}, in Fig. \ref{fig:linder}. In the new ephemeris, the fast rotating secondary is in front of the slowly rotating star at phase 0.0 (which corresponds to phase 0.5 in \citealt{Linder2006}). Although the count-rate light curve does not show obvious evidence of rotational modulation, weak variability with the orbital period is observed, with the lowest counts near the orbital phase $\sim 0.5$. This variation (i.e. maximum during $\phi_{\rm orb} \sim 0.0$, and minimum during $\phi_{\rm orb} \sim 0.5$) is consistent with a colliding-wind binary scenario \citep{Linder2006}.

The results of \citet{Linder2006} must be reconsidered in light of the magnetic field detection by \citet{Grunhut2013}. A standard, pure wind-wind collision between two freely expanding winds is now ruled out, as the Alfvén surface of the broad-lined component extends into the middle of what would otherwise be the collision zone in a non-magnetic system. Depending on the wind momentum balance and magnetic geometry, the primary (narrow-lined) wind may instead interact directly with the magnetically confined plasma of the secondary magnetosphere. Consequently, the X-rays likely form either through magnetically confined wind shocks within the secondary magnetosphere itself \citep[analogous to the prototypical CM system $\sigma$~Ori~E;][]{Townsend2007}, or via a collision between the primary wind and the secondary magnetosphere.

\cite{Kurapati2017} observed the star at 1, 3 and 13 cm radio bands with the Very Large Array (VLA). They reported detection at 1 and 3 cm bands and non-detection at 13 cm band. The authors reported an intra-band spectral index of $-1.2 \pm 1.0$ {between the sub-bands of the X band, which is suggestive of non-thermal emission but does not by itself provide an unambiguous identification of the emission mechanism.}. The small number of observations was insufficient to infer the emission mechanism. The observed radio emission was predicted to be a consequence of three separate components: (i) thermal free–free emission from individual components originating in the surrounding ionized medium, (ii)  nonthermal radio emission from CM of the magnetic secondary, and (iii) additional nonthermal emission from the colliding wind, as expected from the historical interpretation of this system being a colliding wind binary \citep{Wiggs1992}.

In this study, we follow up the enigmatic Plaskett's star, primarily at the X-ray and radio bands. In the X-rays, we report a detailed analysis of X-ray follow-up of this target with the \cxo X-ray telescope. With detailed variability analysis of the X-ray light curve and modeling of high-resolution X-ray spectra, we infer the possible emission mechanisms and origin of emission. To study low-frequency radio emission from this system in the hope of detecting evidence of a colliding wind scenario, we observed the star with the upgraded Giant Metrewave Radio Telescope \citep[uGMRT,][]{Swarup1992, Gupta2017}. We also re-analyzed higher frequency radio data from the VLA, and searched for recent radio surveys.

\begin{deluxetable}{cccccc}
\tablecaption{Observation details of the X-ray observations of the Plaskett's star obtained with \cxo. { The first four observations were taken in January 2016, while the later two were taken in December 2016.} \label{tab:xray_obs}}
\tablehead{
\colhead{ID} & \colhead{Mean HJD} & \multicolumn{2}{c}{Mean Phase} & \colhead{Obs. ID} & \colhead{Exp Time} \\
\colhead{} & \colhead{($-2450000$)} & \colhead{$\phi_{\rm rot}$} & \colhead{$\phi_{\rm orb}$} & \colhead{} & \colhead{(ks)}    
}
\startdata
\multicolumn{6}{c}{{\cxo}} \\
C1 & 2457405.59546 & 0.73 & 0.97 & 17730 & 54.39 \\
C2 & 2457406.38572 & 0.33 & 0.02 & 18751 & 43.97 \\
C3 & 2457408.67453 & 0.22 & 0.18 & 17731 & 46.49 \\
C4 & 2457410.48697 & 0.75 & 0.31 & 18752 & 55.38 \\
C5 & 2457728.50629 & 0.35 & 0.40 & 19955 & 46.99 \\
C6 & 2457730.19733 & 0.86 & 0.53 & 17732 & 72.65 \\
\enddata
\end{deluxetable}

This paper is structured as follows. The observation details and basic data reduction procedures are explained in Section \ref{sec:Observations}. In Sections \ref{sec:xray_lc}, \ref{sec:xray_spec_low}, and \ref{sec:xray_spec_high}, we present the X-ray study, discussing the results from the X-ray light curve, lower resolution ACIS spectra, and high resolution HETG spectra, respectively. In Section \ref{sec:radio}, we discuss the radio observations. In section \ref{sec:Discussion_Conclusion}, we summarize the multi-wavelength results and their possible interpretations.

\section{Observation and Data Reduction} \label{sec:Observations}

\subsection{\cxo}

\begin{figure}
    \centering
    \includegraphics[width=\linewidth]{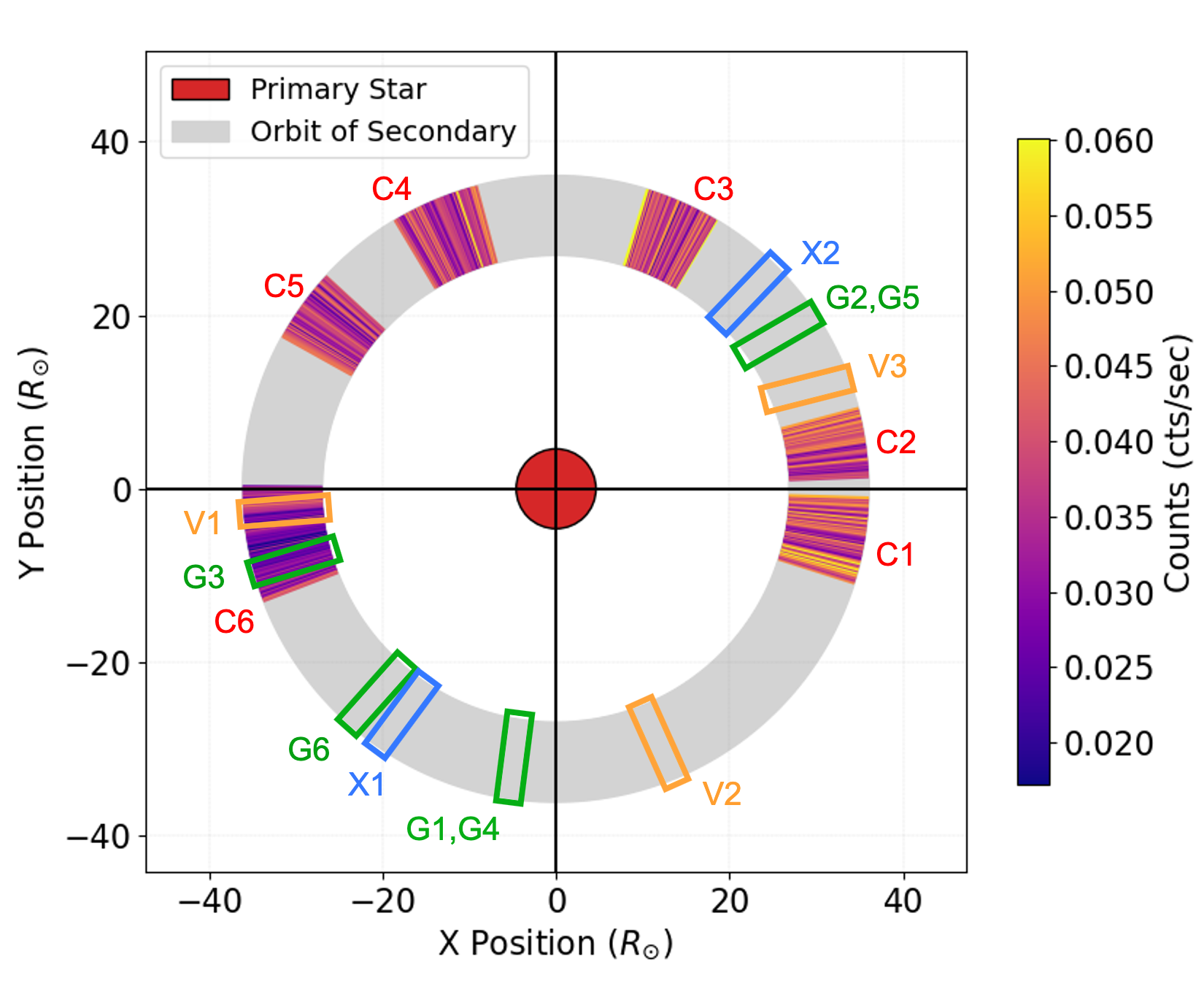}
    \caption{Representation of the X-ray and radio observations along the orbit of Plaskett's star (not to scale). The narrow colored bands represent the position of \cxo observations reported in this paper. The color scale shows the count rate from \cxo observations at each phase, binned every 15 minutes. The observation IDs from Table \ref{tab:xray_obs} are indicated around the circle. The uGMRT phases are shown with green boxes, the VLA observations with orange boxes, and the \xmm observations with blue boxes. An observer from earth will be at position $+\infty$ of the x-axis. { $\phi_{\rm orb}=0$ corresponds to the secondary in front, on the positive X-axis.}}
    \label{fig:orbit}
\end{figure}

X-ray spectroscopic observations of the Plaskett's star with \cxo were taken between 8 January 2016 and 7 December 2016 under project 17200115 (PI: M. Leutenegger). The observations were taken at six epochs, each epoch covering almost half rotational period (thus covering 3 full cycles in total). The observations also take place at 6 distinct orbital phases. The ranges of the observations are represented in Fig. \ref{fig:orbit}. The observation details can be found in Table \ref{tab:xray_obs}. In
all observations, the \cxo ACIS-S camera was operated with the High Energy Transmission Grating (HETG) in the Timed Exposure (TE) read mode and the VFAINT data mode.

The data were reduced using the CIAO package {(version 4.16)} and the Calibration Database {(CALDB v4.11)}. Calibrated event files were obtained with the CIAO task \textit{`chandra\textunderscore repro'}. All X-ray spectral analysis reported in this paper were performed with the \textit{xspec\footnote{\url{https://heasarc.gsfc.nasa.gov/xanadu/xspec/}}} package v.12.13.1 \citep{Arnaud1996}. A python wrapper \textit{pyxspec\footnote{\url{https://heasarc.gsfc.nasa.gov/xanadu/xspec/python/html/index.html}}} was utilized for automated analysis of a large number of spectra files. More details on the light curve and spectral analysis are given in the proceeding sections.

The target was also observed with \xmm twice under proposal 000173 (PI: G. Rauw), the first on 17 September 2002 and the second on 16 March 2003. Coincidentally, both observations were taken during the same rotation phase (see Table \ref{tab:xray_obs}). The results from these observations were originally published in \cite{Linder2006}, and are re-analyzed here with similar procedure. We used the XMM–Newton Science Analysis System (SAS, Version: 21.0.0)\footnote{\url{https://www.cosmos.esa.int/web/xmm-newton/sas}} for data reduction. Event files filtered using Good Time Intervals (GTI) was created from the SAS task \textit{`xmmextractor'}.  This GTI-filtered event file was further used to generate custom spectra and light curves. These data were used only for the spectral analysis that is presented in the Appendix \ref{App:spectra}.

\subsection{uGMRT}

Given the possibility of detecting nonthermal emission triggered by colliding wind from this binary system, as predicted by \cite{Kurapati2017}, we observed the target with uGMRT uniformly covering the orbital period of the star. To obtain reliable interband spectra at different phases, we performed simultaneous measurements in band 4 (550--900 MHz) and band 5 (1050--1450 MHz)  in dual subarray mode, with a nearly uniform distribution of antennas between these two arrays. The observations were taken at six different epochs between 26 July 2020 and 5 September 2020 (see Table \ref{tab:radio_obs}) during uGMRT Cycle 38 (ObsID: 38\textunderscore 102, PI: P. Chandra). We excluded the latest observation on 5th September due to high noise level. Each observation spanned approximately 4 to 5 hours, with a typical on-source time of 3 hours. Both bands were divided into 2048 channels and the observations were taken in full polar mode. 3C147 (J0542+498) was used as the amplitude calibrator, and J0745+101 was used as the phase calibrator during all observations.

\begin{deluxetable}{llcccc}
\tablecaption{Observation details and flux densities at different epochs from the uGMRT observations (top), VLA (middle), and Rapid ASKAP Continuum Survey observations (bottom). {For uGMRT observations, B4 denotes band 4 (550--900 MHz) and B5 denotes band 5 (1050--1450 MHz) observations.} $\sigma$ denotes the noise level near the target, and $F_\nu$ represents the flux density (or 3-$\sigma$ upper limits in case of nondetections) in mJy. \label{tab:radio_obs}}
\tablehead{
\colhead{Mean HJD}  & \multicolumn{2}{c}{Mean Phase} & \colhead{Freq. Band} & \colhead{$\sigma$}  & \colhead{$F_{\nu}$}  \\
\colhead{ } & \colhead{$\phi_{\rm rot}$} & \colhead{$\phi_{\rm orb}$} & \colhead{} & \colhead{(mJy)} & \colhead{(mJy)}
}
\startdata
\multicolumn{6}{c}{uGMRT (Targeted)} \\
2459057.81 & 0.74 & 0.72 & B5 &   0.032 & $<0.096$ \\
 &  &  & B4  & 0.039 & $<0.117$ \\
2459063.70 & 0.59 & 0.13 & B5 & 0.021 & $<0.063$ \\
 &  &  & B4 & 0.034 & $<0.102$ \\
2459069.63 & 0.47 & 0.54 & B5 & 0.028 & $<0.084$ \\
 &  &  & B4 & 0.70 & $<0.210$ \\
2459086.69 & 0.51 & 0.72 & B5 & 0.027 & $<0.081$ \\
 &  &  & B4 & 0.026 & $<0.078$ \\
2459092.56 & 0.33 & 0.13 & B5 & 0.022 & $<0.067$ \\
 &  &  & B4 & 0.047 & $<0.141$ \\
\multicolumn{6}{c}{VLA (Re-analysed)} \\
2456852.13 & 0.14  & 0.51 & Ka (32 GHz) &  0.016 & $0.40 \pm 0.04$  \\
2456784.47 & 0.47 &  0.80 & X (10 GHz)  &  0.008 & $0.21 \pm 0.02$  \\
2456851.95 & 0.71  & 0.08  &S (3.2 GHz) &  0.007 & $0.06 \pm 0.02$  \\
\multicolumn{6}{c}{ASKAP} \\
2459585.19  & 0.62 & 0.35  & 1655.5 MHz  &  0.34 & $<1.02$  \\
2459582.13  & 0.11 & 0.14  & 1655.5 MHz  &  0.28 & $<0.84$  \\
2459233.11  & 0.96 & 0.89  & 1367.5 MHz  &  0.27 & $<0.81$  \\ 
2459232.12  & 0.15 & 0.82  & 1367.5 MHz  &  0.25 & $<0.75$  \\
2460308.21  & 0.45 & 0.57  & 943.5 MHz   &  0.23 & $<0.69$  \\
2460307.22  & 0.64 & 0.50  & 943.5 MHz   &  0.23 & $<0.69$  \\
2458598.77  & 0.09 & 0.83  & 887.5 MHz   &  0.33 & $<0.99$  \\
2458598.77  & 0.34 & 0.77  & 887.5 MHz   &  0.46 & $<1.38$  \\
\enddata
\end{deluxetable}

All data were analyzed using the Common Astronomy Software Applications (CASA\footnote{\url{https://casa.nrao.edu/}}) package \citep{McMullin2007}. We used the calibration and imaging process described in \cite{Biswas2023}. The calibrated data for the target were averaged over four frequency channels. The approximate usable bandwidth is 370 MHz for Band 5, and $\sim$290 MHz for Band 4. The imaging was performed with the CASA task `{\it tclean}' along with the `{\it mtmfs}' deconvolver (Multiscale Multifrequency with W-projection, \citealt{Rau2011}) using both interactive masking and the automated `\textit{auto-multithresh}' masking. We also removed strong sources using the task `\textit{uvsub}' to gain better sensitivity. In addition, several rounds of self-calibration were performed for better imaging quality.

\subsection{VLA and ASKAP}

The target was observed with VLA during a survey project (Project ID 14A-139, PI: P. Chandra, \citealt{Kurapati2017}) in the S (2--4 GHz), X (8--12 GHz), and Ka (29--37 GHz) bands between May and July 2014. We re-analyzed the results originally published by \cite{Kurapati2017}. The initial flagging was performed with CASA, and the calibration process was performed with the VLA calibration pipeline built into the CASA pipeline version. Further imaging processes are similar to the uGMRT data mentioned in the previous subsection. We detected the target in all bands, including the S band, which was previously reported as non-detection in \cite{Kurapati2017}. {The S-band observation was taken with VLA A-configuration and resulted in a synthesized beam size of $0.66" \times 0.56"$. The Ka and X band observations were taken in D-configuration. The Ka band {synthesized} beam size was $2.69" \times 1.81"$. The target was unresolved in all bands}. The updated flux densities are shown in Table. \ref{tab:radio_obs}.

We searched for radio emission from the star from the Rapid ASKAP Continuum Survey (RACS; \citealt{McConnell2020, Hale2021, Duchesne2023}). RACS is a continuum survey using the Australian SKA Pathfinder (ASKAP), obtained at three frequency bands (RACS-low: 887.5 MHz, RACS-mid: 1367.5 MHz, and RACS-high: 1655.5 MHz). The target was not detected in any band. The upper limits from the RACS are also shown in Table. \ref{tab:radio_obs}.

\subsection{\textit{TESS}}

In order to compare the rotational modulation from emission at different frequency bands (or magnetic measurements), we also report the photometric observations of Plaskett's star taken with the Transiting Exoplanet Survey Satellite (\textit{TESS}) space telescope \citep{Ricker2014}. \textit{TESS} divides its observation time into 26 partially overlapping 24$^o \times 96^o$ sectors, and during each year, \textit{TESS} observes 13 sectors for 27 d each. It operates in the wavelength range 600-1050 nm. Our target was observed only once by \textit{TESS} as part of sector 6 from December 15, 2018 to January 6, 2019.

\begin{figure}
    \centering
    \includegraphics[width=0.95\linewidth]{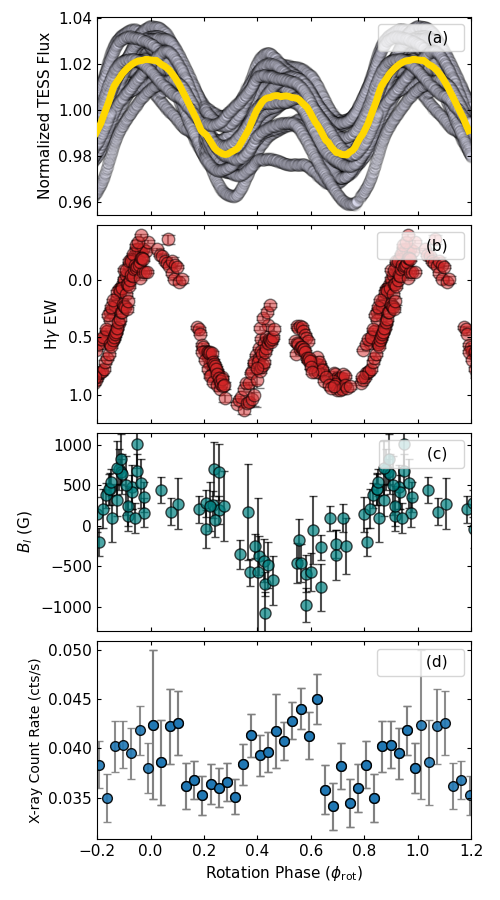}
    \caption{Rotational modulation shown by Plaskett's star in different forms: (a) The photometry observation obtained with \textit{TESS}. The grey points show the cycle-to-cycle variability, while the yellow curve shows the binned light curve. All the light curves are binned with $\Delta \phi = 0.01$. (b) Reproduction of the H$\gamma$ equivalent width variation from the star obtained from ESPaDOnS observation from \cite{Grunhut2022}. (c) The longitudinal magnetic field variation with rotation phase, with data obatained from \cite{Grunhut2022}. (d) The X-ray light curve binned every $\sim 0.3$ phase with \cxo. More details on the X-ray light curve can be found in the text. All the light curves are phased according to the ephemeris reported by \cite{Grunhut2022}.}
    \label{fig:compare_lc}
\end{figure}

The processed 2-minute cadence light curves were obtained from the Mikulski Archive for Space Telescopes (MAST, \url{https://archive.stsci.edu/}). The periodogram of the light curve shows the first harmonic of the orbital period to be the strongest frequency, whereas the rotation period is found to be the second strongest. We performed a Markov Chain Monte Carlo (MCMC) analysis using two periods. The best fit periods are found to be $\approx0.60757$ days and $\approx1.21557$ days, respectively. These periods are consistent with the rotation period obtained by \cite{Grunhut2022} from the equivalent width variation and the longitudinal magnetic field variation, within uncertainty. However, we noticed a slight mismatch of the reference HJD. A phase shift of $+0.15$ was required to properly phase the data according to the ephemeris reported in \cite{Grunhut2022}, which may have resulted from the accumulated error in the ephemeris. The phase folded \textit{TESS} light curve is shown in Fig. \ref{fig:compare_lc}a, along with the overall binned light curve to emphasize overall periodicity. The light curve shows significant cycle-to-cycle variability. The overall double wave pattern is similar to the H$\gamma$ effective width variation (Fig. \ref{fig:compare_lc}b).

\section{X-ray Light Curve}  \label{sec:xray_lc}

The \cxo observations uniformly covered 3 rotation periods in the 6 observing runs, covering nearly 315 ks. For each \cxo observation, we extracted unique source and background regions to avoid contamination due to a nearby source or detector boundary. In the case of the \xmm data, \cite{Linder2006} noticed that some nearby sources potentially contaminate the target. However, due to its higher resolution, these sources are not contaminating the \cxo data. In this section, we focus only on the \cxo light curve from the ACIS dataset. The light curves were extracted from the event files with the CIAO task \textit{`dmextract'}. For all light curves, barycentric correction was performed and background flares were removed. Fig. \ref{fig:orbit} shows the variation of the X-ray count rate throughout the orbit, where each band represents a 15-minute binned count rate. We did not find any significant short-term variability, as also found in the \xmm light curve \citep{Linder2006}.

\begin{figure}
     \centering
     \begin{subfigure}[b]{0.42\textwidth}
         \centering
         \includegraphics[width=\textwidth]{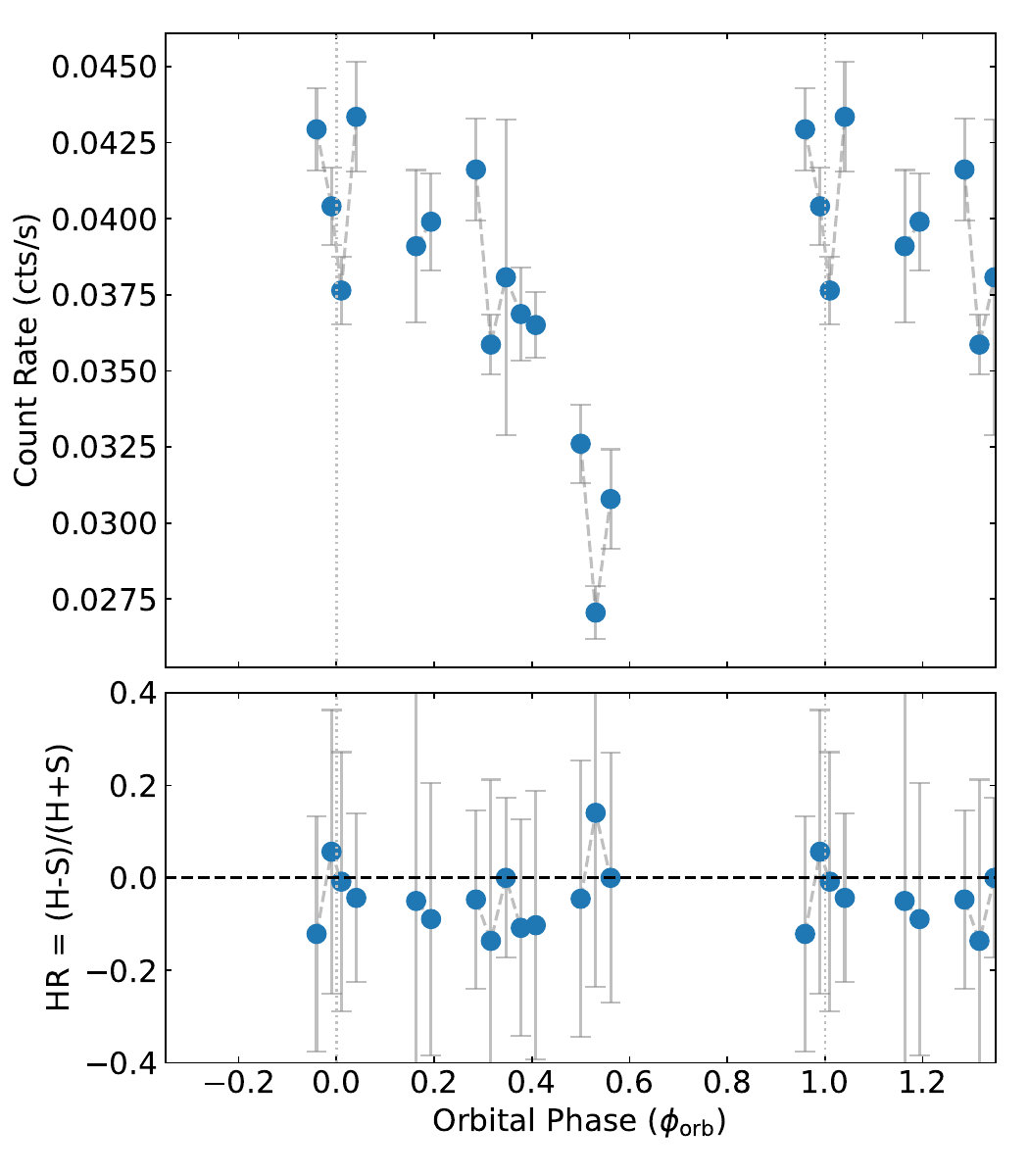}
         \caption{Orbital variation of X-ray count rate and HR}
         \label{fig:x_ray_orb1}
     \end{subfigure}
     \begin{subfigure}[b]{0.42\textwidth}
         \centering
         \includegraphics[width=\textwidth]{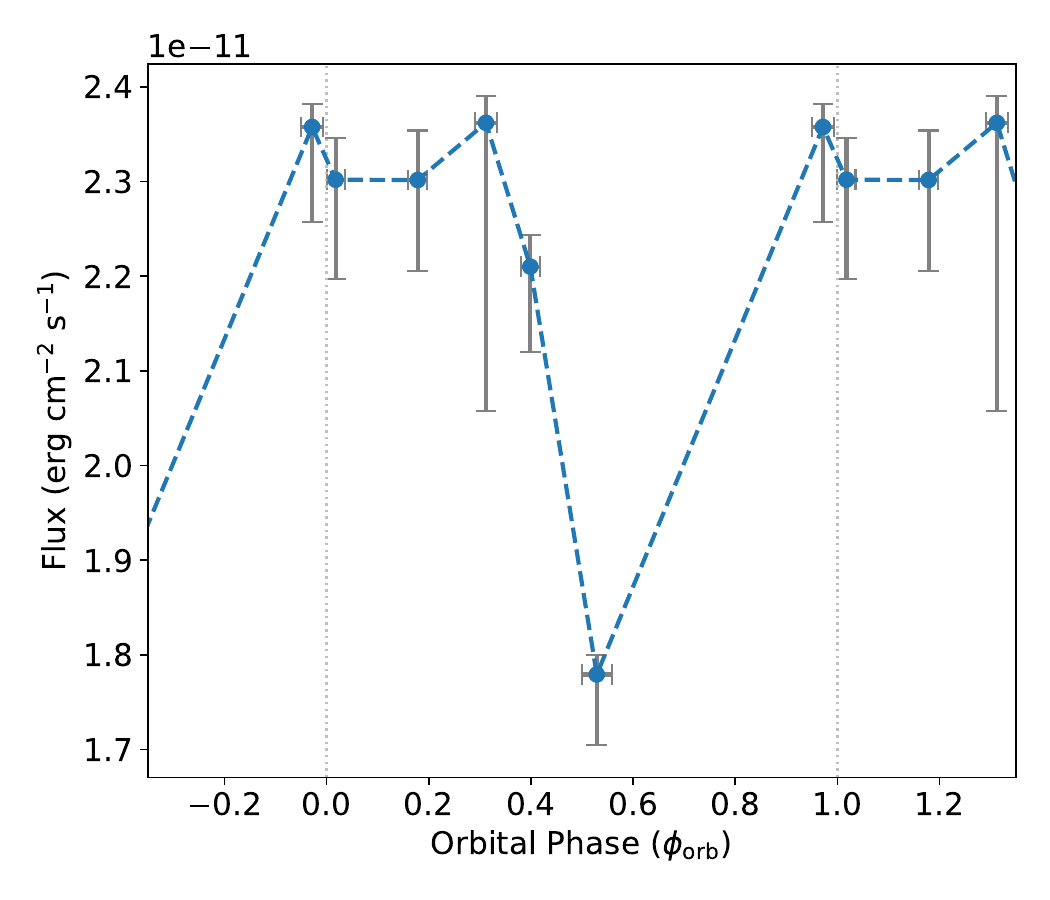}
         \caption{Orbital variation of X-ray flux}
         \label{fig:x_ray_orb2}
     \end{subfigure}
        \caption{(a) \cxo count rate light curves (top panels) and hardness ratio variation (bottom panels) binned and phased according to orbital period. The light curves are binned every $\Delta \phi \sim 0.03$. The hardness ratio (HR) is defined as HR = (H-S)/(H+S), where H is the hard X-ray count (2--10 keV), and S is the soft X-ray count (0.2--2 keV). (b) Orbital flux variation obtained from spectral fitting in the enrgy range 0.5--10 keV. \label{fig:x_ray_orb}}
\end{figure}

\subsection{Orbital Modulation}  \label{OrbMod}

The \cxo observations cover primarily the first half of the orbital period (phases 0.95 to 0.55). We noticed a gradual decrease of the X-ray count rate toward phase 0.5, i.e. stellar conjunction with the slow rotating primary in front of its companion (Fig. \ref{fig:x_ray_orb}). The decrease in count rate is significant ($>30$\%) {and similar to the trend observed by \cite{Linder2006} (Fig. \ref{fig:linder}, bottom)}.

{
Such a variation of thermal X-rays is indeed expected from a massive binary with winds colliding at their terminal velocities \citep{Pittard1997, Pittard2010}. For the secondary star, following the mass-loss calculation formalism of \cite{Vink2021} and \cite{Vink2001}\footnote{\url{https://www.astro.physik.uni-potsdam.de/~ansander/mass-loss-recipes/mdotvink.py}} and the parameters derived by \cite{Wade2026}, we determine the value of $\dot{M}_2$ to be $\sim 1.2 \times 10^{-7}\ M_{\odot}\,{\rm yr}^{-1}$. For the primary, we derive $\dot{M}_1 \sim 1.2 \times 10^{-6}\ M_{\odot}\,{\rm yr}^{-1}$. These should be regarded as adopted theoretical estimates rather than direct measurements. Independently, \cite{Wade2026} obtain $\log(\dot{M}_1/M_{\odot}\,{\rm yr}^{-1})=-5.8\pm0.3$ from their wind analysis, corresponding to $\dot{M}_1\simeq1.6\times10^{-6}\,M_{\odot}\,{\rm yr}^{-1}$, broadly consistent with our adopted value. Their analysis also emphasizes systematic uncertainties in modeling the primary wind, while the secondary wind is particularly difficult to constrain because of contamination by the magnetosphere. \cite{Pittard1997} computed synthetic X-ray light curves for different wind momentum ratios, which can be expressed in terms of

\begin{equation}
\mathcal{R} = \left( \frac{\dot{M}_1 v_{\infty,1}}{\dot{M}_2 v_{\infty,2}} \right)^{1/2},
\end{equation}

\noindent
where $\dot{M}_1$, $\dot{M}_2$ are the mass-loss rates of the primary with the stronger wind and the secondary with the weaker wind, and $v_{\infty,1}$, $v_{\infty,2}$ are the corresponding wind terminal velocities. { For purposes of an initial estimate of the wind momentum balance ratio we adopt the simplifying assumption that the winds are freely expanding and have similar terminal velocities.} This results in $\mathcal{R} \sim 3.2$, assuming similar wind-terminal velocities. For this value of $\mathcal{R}$, one can expect a minimum flux near orbital phase 0.5, i.e., at phases when the star with the stronger wind (in our case the non-magnetic primary) is in front. A similar explanation was also provided by \cite{Linder2006}, but the authors were forced to provide additional considerations due to a different calculation of mass-loss.

\begin{figure*}
\centering
\gridline{\fig{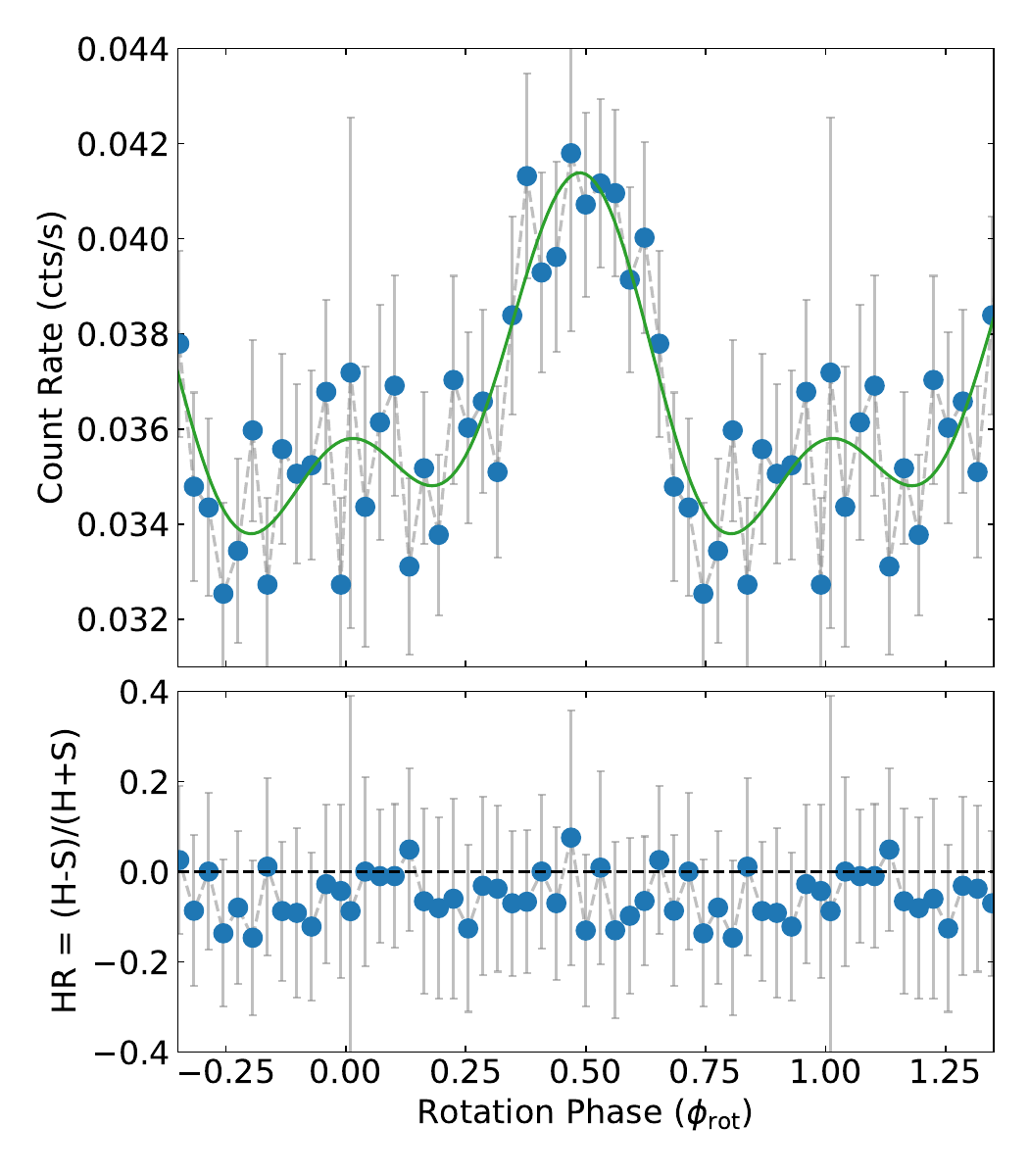}{0.38\textwidth}{(a) Rotational Variation ($P_{\rm rot} \sim 1.22$ d)} \label{fig:x_ray_rot}
          \fig{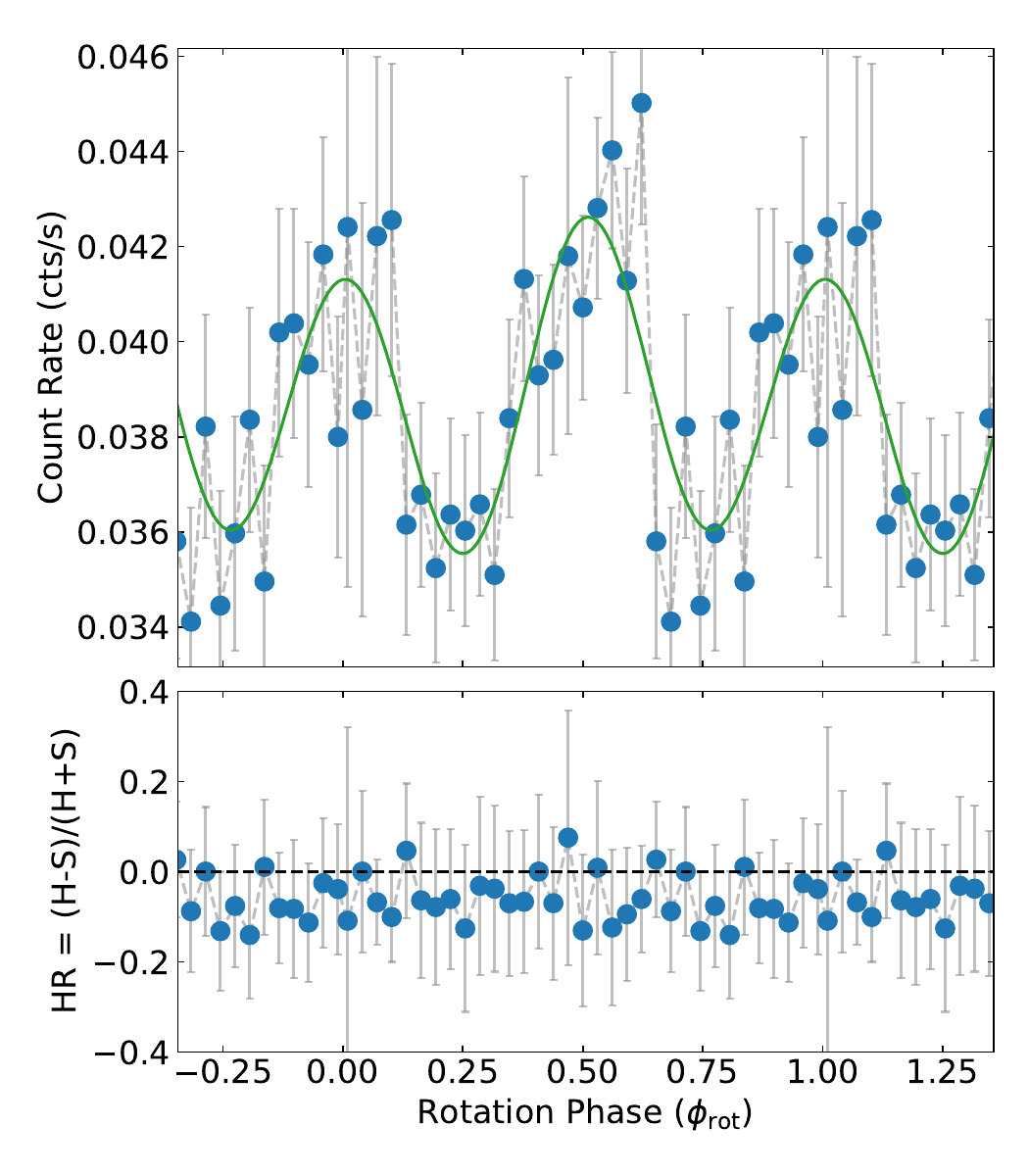}{0.38\textwidth}{(b) Excluding Obs. C6} \label{fig:x_ray_rot_b}}
\gridline{\fig{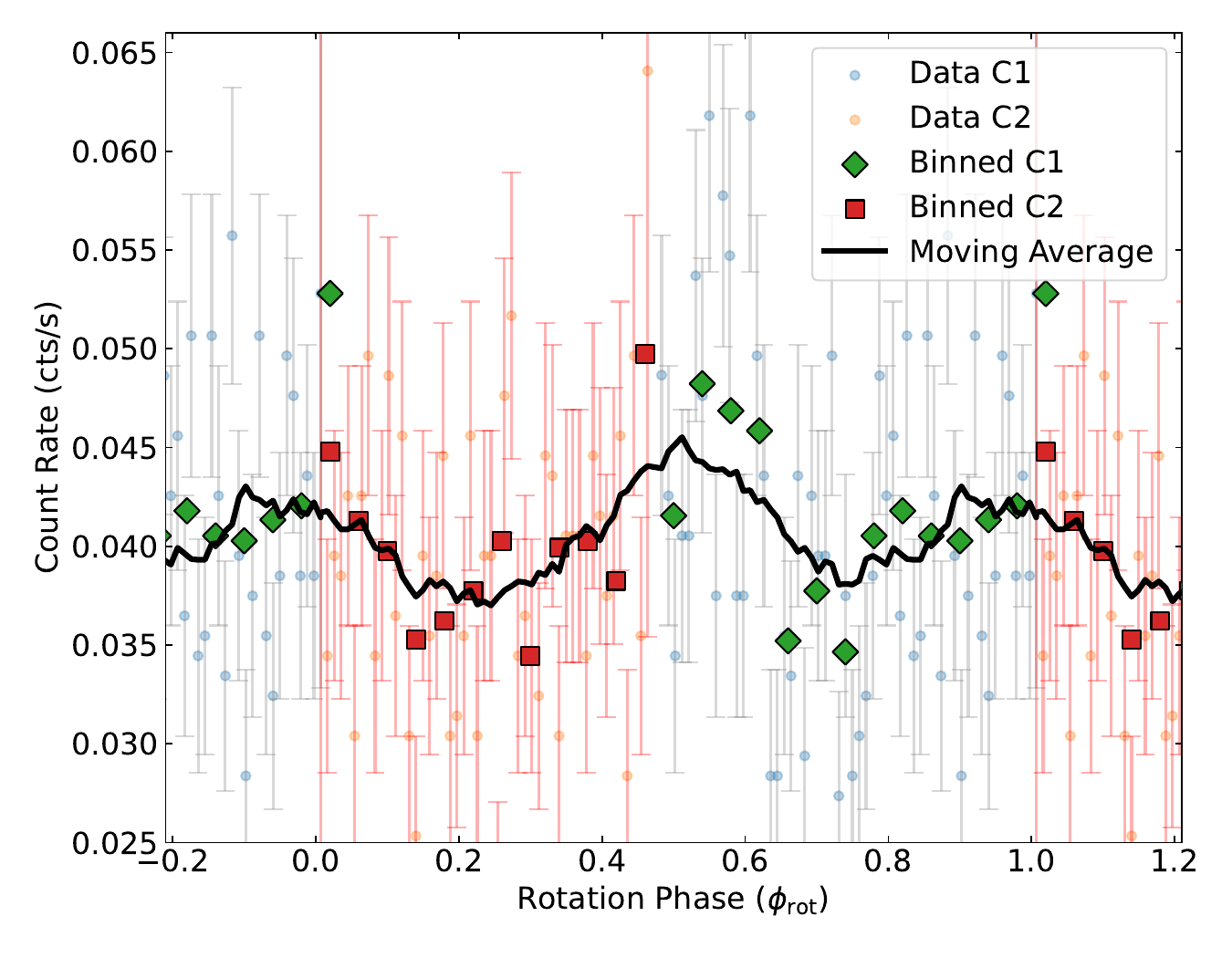}{0.42\textwidth}{(c) C1 and C2 observations that cover similar orbital phase but different rotation phase windows.} \label{fig:x_ray_C1C2}
        \fig{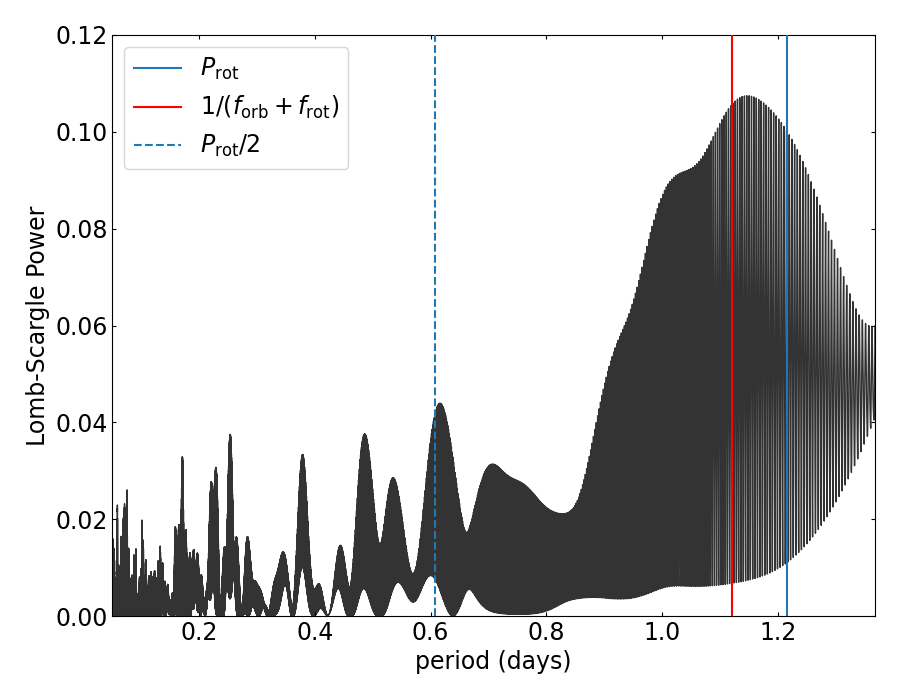}{0.42\textwidth}{(d) Periodogram (all observations)} \label{fig:x_ray_periodo}}
    \caption{\textbf{(a \& b):} \cxo count rate light curves (top panels) and hardness ratio variation (bottom panels): binned and phased according to rotation period (a) with all the observations, and (b) with every observations except C6 (orbital phase $\sim 0.5$), respectively. In both cases, the light curves are binned every $\Delta \phi \sim 0.03$. \textbf{(c)} The light curve from C1 and C2, which correspond to similar orbital phase ($\sim 0.0)$ but spans different rotational phases, and \textbf{(d)} Lomb-Scargle periodogram obtained from combining all 100-sec light curves. \textbf{(e)} }
    \label{fig:chandra_lc}
\end{figure*}

The above calculation provides a useful reference for the standard colliding-wind interpretation. However, the magnetic field of the secondary requires the geometry of the interaction to be considered explicitly. In the standard ram-pressure balance of two freely expanding winds, the distances of the shock point from the primary and secondary are given by

\begin{equation}
r_1 = \frac{\mathcal{R}}{1+\mathcal{R}}a,
\qquad
r_2 = \frac{1}{1+\mathcal{R}}a,
\end{equation}

\noindent
where $r_1+r_2=a$ and $\mathcal{R}=(\dot{M}_1 v_{\infty,1}/\dot{M}_2 v_{\infty,2})^{1/2}$  \citep{Stevens1992,Pittard1997}. Following \cite{Wade2026}, we adopt an orbital separation of $a=78.4^{+5.5}_{-4.5}\,R_{\odot}$. For $\mathcal{R}\simeq3.2$, this gives $r_1\simeq60,R_{\odot}$ from the primary and $r_2\simeq19,R_{\odot}$ from the secondary. With the secondary radius of $R_2\simeq10.4,R_{\odot}$ inferred by \cite{Wade2026}, the nominal interaction point therefore lies at only $\sim1.8R_2$ from the secondary.

This result is particularly important in the context of the magnetic secondary. The Alfv{\'{e}}n radius of the secondary is $R_{\rm A}\simeq4.4R_2$ \citep{Grunhut2022}. Thus, the nominal shock location obtained from the non-magnetic wind-momentum balance lies well inside the magnetic confinement region of the secondary. Consequently, the above calculation should not be interpreted as demonstrating that the two stellar winds collide at this location. Rather, it shows that, if the winds were treated as freely expanding {with equal terminal velocities}, their nominal interaction point would occur deep within the magnetosphere of the secondary. The primary wind is therefore expected to interact with the magnetically confined wind and closed-field region of the secondary star rather than with an independently expanding secondary wind. The actual location and geometry of the interaction will depend on the three-dimensional magnetic field and wind structure. This treatment is beyond the scope of this study. In this picture, the observed orbital modulation can still be understood naturally as a phase-dependent change in the visibility and attenuation of the X-ray-emitting plasma associated with the interaction region. In particular, the decline toward phase 0.5, when the stronger-wind primary lies in front, is qualitatively consistent with the increased attenuation by the primary wind, as found in standard colliding-wind calculations \citep{Pittard1997}. }

\subsection{Rotational Modulation}

We investigated the rotational modulation of the X-ray light curve by binning the folded light curve every $\Delta \phi_{\rm rot} \sim 0.03$ { using $P_{\rm rot}=1.21551$ d. The rotational reference phase corresponds to the photometric maximum adopted from \cite{Grunhut2022}, at which the north magnetic hemisphere of the secondary is visible}.  The resulting light curve is shown in Fig. \ref{fig:chandra_lc}a. We noticed obvious variability in the X-ray count rate on rotational timescales ($\sim 20$\%), with the maximum count near the phase $\phi_{\rm rot} \sim 0.5$. The X-ray light curve is not a proper sinusoid but can be fitted well with the rotation period and its first harmonic (green curve in Fig. \ref{fig:chandra_lc}a). This behavior is similar to the variation of H$\alpha$ or H$\beta$ emission from this star. However, optical photometry (and H$\gamma$ equivalent width variation) shows a higher peak at phase 0.0 and a smaller second peak near phase 0.5 (Fig. \ref{fig:compare_lc}). In order to exclude any bias contributed by the binary variation at phase 0.5, we repeated the same analysis excluding the corresponding observation (C6). The resulting light curve is shown in Fig. \ref{fig:chandra_lc}b. This variation is similar to the photometric variation and equivalent width variation of H$\gamma$ shown in Fig. \ref{fig:compare_lc}, { and provides strong evidence that a substantial fraction of the X-ray emission is associated with the magnetosphere of the secondary star.}

We also examined two observations with similar orbital phase (C1 and C2 with $\phi_{\rm orb} \sim 0.0$) with different rotational phase coverage in Fig. \ref{fig:chandra_lc}c. We reproduce a light curve similar to that in Fig. \ref{fig:chandra_lc}a,b. This treatment also demonstrates that the rotational variation is independent of the orbital variation. Note that the variation is not immediately evident in the spectra from individual observations. In Appendix \ref{App:synthetic}, we demonstrate that, given the instrumental uncertainty, the binning of the spectra is necessary to distinguish the variation with high significance. We also show in Appendix \ref{App:synthetic} that the orbital variation cannot randomly produce any such rotational variation. We also searched for variation of the hardness ratio (HR). Here, we define the hardness ratio as HR = (H-S)/(H+S), where H is the hard X-ray count (2--10 keV) and S is the soft X-ray count (0.2--2 keV). The observed HR light curve is moderately soft, and does not show any significant variation with either rotational or orbital phase.

\begin{deluxetable*}{cccccccccc}
\tablecaption{The `2T' spectral fitting results obtained with \textit{xspec} from \cxo data. The $n_{\rm H}$ reported in the second column correspond to the additional absorption model, rather than the fixed ISM absorption. The numbers in parenthesis in the $\chi^2_{\rm reduced}$ column represents the number of degrees of freedom. T$_{\rm apec}$ represents the best-fit temperatures from \textit{apec} models. The normalization parameters are in units of $10^{-3}$ cm$^{-5}$.  All spectra were fitted in the energy range 0.5--10.0 keV. The interstellar medium (ISM) contribution for this system is known to be small ($0.67 \times 10^{21}$ cm$^{-2}$). The flux values represented here are the unabsorbed flux after ISM correction.  The complete model expression used in \textit{xspec} is: `\textit{tbabs*phabs*(apec+apec)}'.    \label{tab:2T_fit}} 
\tablehead{
\colhead{Data} & \colhead{$\phi_{\rm orb}$} & \colhead{$n_{\rm H}$} & \colhead{$\chi^2_{\rm red}$}  &  \multicolumn{2}{c}{T$_{\rm apec}$ (keV)} & \multicolumn{2}{c}{Norm ($\times 10^{-3}$ cm$^{-5}$)} & \colhead{$\log F_{(\rm 0.5-10 \ keV)}$} & \colhead{HR}  \\
\colhead{ID}  & &  \colhead{$\times 10^{22}$ cm$^{-2}$} & \colhead{} & \colhead{kT$_1$} & \colhead{kT$_2$} & \colhead{$N_{\rm kT1}$} & \colhead{$N_{\rm kT2}$} &  \colhead{(erg cm$^{-2}$ s$^{-1}$)} & \colhead{(H-S)/(H+S)}
}
\startdata
C1 & 0.95 &  0.23 $\pm$ 0.07 & 1.23(112) & 0.73 $\pm$ 0.03 & 2.45 $\pm$ 0.15 & 1.0 $\pm$ 0.3 & 1.3 $\pm$ 0.1 & -11.514 $\pm$ 0.011 & -0.40 $\pm$ 0.02  \\
C2 & 0.00 & 0.10 $\pm$ 0.07 & 1.33(84) & 0.88 $\pm$ 0.04 & 2.30 $\pm$ 0.12 & 0.6 $\pm$ 0.2 & 1.4 $\pm$ 0.1 & -11.528 $\pm$ 0.013 & -0.39 $\pm$ 0.02      \\
C3 & 0.16 & 0.24 $\pm$ 0.07 & 1.29(87) & 0.77 $\pm$ 0.03 & 2.31 $\pm$ 0.15 & 1.0 $\pm$ 0.3 & 1.3 $\pm$ 0.1 & -11.528 $\pm$ 0.012 & -0.41 $\pm$ 0.02    \\
C4 & 0.29 & 0.34 $\pm$ 0.09 & 1.13(97) & 0.33 $\pm$ 0.04 & 1.96 $\pm$ 0.09 & 3.0 $\pm$ 1.5 & 1.8 $\pm$ 0.1 & -11.499 $\pm$ 0.012 & -0.44 $\pm$ 0.02  \\
C5 & 0.38 & 0.19 $\pm$ 0.08 & 1.28(85) & 0.83 $\pm$ 0.04 & 2.31 $\pm$ 0.14 & 0.8 $\pm$ 0.2 & 1.3 $\pm$ 0.1 & -11.548 $\pm$ 0.013 & -0.39 $\pm$ 0.02  \\
C6 & 0.50 & 0.00 $\pm$ 0.07 & 1.48(107) & 0.88 $\pm$ 0.05 & 2.59 $\pm$ 0.14 & 0.3 $\pm$ 0.1 & 1.1 $\pm$ 0.1 & -11.625 $\pm$ 0.013 & -0.35 $\pm$ 0.02   \\
\enddata
\end{deluxetable*}

\begin{figure}
    \centering
    \includegraphics[width=0.47\textwidth]{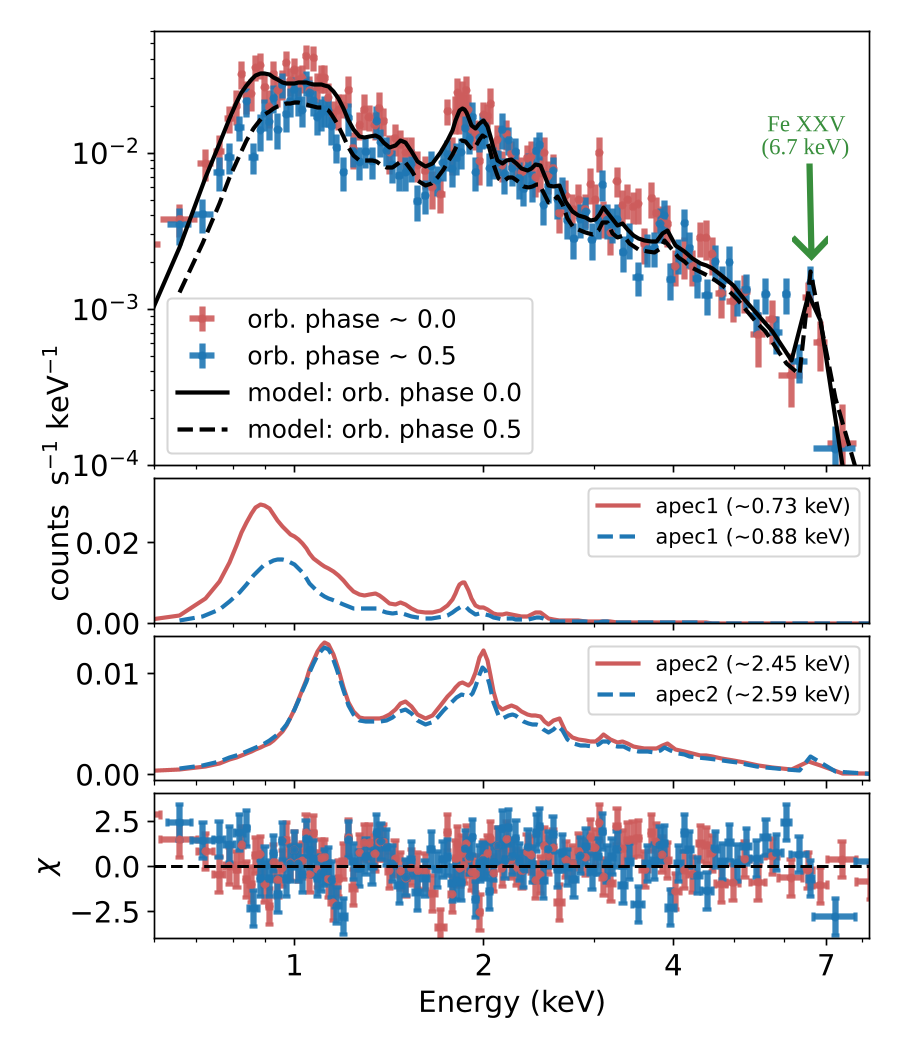}
    \caption{Comparison of \cxo ACIS spectra during orbital phase 0.0 (in red) and 0.5 (blue). The top panel shows the binned spectra, with the best fit model for phase 0.0 (solid black curve) and phase 0.5 (dashed back curve), respectively. The middle two panels show the contribution from the model components in each case. The bottom panel shows the residual as $\chi$, where it is defined as $\chi$ = (data-model)/error. {The Fe XXV line at 6.7 keV is shown with a green arrow.} }
    \label{fig:example_spectra}
\end{figure}

A correlation of optical and X-ray light curve on rotational time scale was previously seen in some magnetic OB stars (e.g. $\theta^1$ Ori C, \citealt{Gagne2005}, Tr16-22, \citealt{Naze2014a}, NGC1624-2, \citealt{Petit2015}, CPD --28$^o$ 2561, \citealt{Naze2015}; HD 108, \citealt{Rauw2023}, HD 191612, \citealt{Naze2010}). In the case of an edge-on system such as NGC1624-2, absorption can be the origin of such variation where additional absorption by the dense magnetosphere can reduce the soft X-ray flux \citep{Petit2015}. In the tilted dipole geometry, the variation of X-ray flux is expected due to the regular occultation of the X-ray emitting region by the stellar disk \citep{ud-Doula2016}. The $\sim 48^o$ inclination and $\sim 90^o$ obliquity make Plaskett's star a good candidate for rotational variation, thus explaining the observed variation. However, it should be noted that the large CM of the star can affect occultation and modify the variability \citep{Naze2014}. {This is because, in all of the cases mentioned above, the stars are assumed to host dynamical magnetospheres (DMs), in which the confined wind plasma is not centrifugally supported and can therefore be approximated as a disk or torus around the magnetic equator}. However, for a CM, the rotation is dynamically significant and distorts the magnetosphere \citep{Townsend2005}. 

\section{Lower Resolution X-ray Spectra} \label{sec:xray_spec_low}

We searched for variation in the X-ray spectra from the \cxo ACIS data. Spectra from all six observations were fitted with two multiplicative absorption models and multiple optically thick thermal plasma models (\textit{apec}). One of the absorption models (\textit{tbabs}) was kept fixed to take into account the ISM absorption ($0.67 \times 10^{21}$ cm$^{-2}$, \citealt{Linder2006}). Another absorption model (\textit{phabs}) was used to take into account any additional circumstellar absorption. We also investigated a different model with power-law components in search of nonthermal X-ray produced by wind-wind collision, producing a possible inverse Compton tail. Thus, we considered three model variations: (i) 2-temperature thermal (`2T')\footnote{The complete model expression used in \textit{xspec} for this case is: `\textit{tbabs*phabs*(apec+apec)}'.}, (ii) 3-temperature thermal (`3T'), and (iii) 2-temperature thermal with a power-law component (`2T+pow').

We show the spectra from two different orbital phases in Fig. \ref{fig:example_spectra}. {The spectra appear to be thermal; an Fe~{\sc xxv} feature near 6.7 keV is visible in some of the Chandra ACIS spectra. The feature is clearly apparent in observations C1, C2, and C4, but is less evident in C3 and C5. The intermittent visibility of this feature may be related to the signal-to-noise of the individual observations, although intrinsic or orbital variability cannot be excluded. A detailed investigation of the Fe~{\sc xxv} line variability would require higher signal-to-noise spectra and is beyond the scope of the present work.} A fit to the data with the `2T' model, `3T' model and `2T+pow' model gives a similar quality of fit.  However, given the thermal nature and quality of fit, we prefer the simplest `2T' model. The best fit parameters with `2T' model obtained with \textit{xspec}  are shown in Table \ref{tab:2T_fit}. The best-fit parameters obtained from other models and individual spectra with their corresponding model fits are discussed in Appendix \ref{App:spectra}.  It should be noted that the EPIC-pn data from the second short \xmm observation gives a poor fit to both the `2T' and `3T' models, compared to the power-law model \cite{Linder2006}. However, given the short duration of this \xmm dataset, we excluded it from the general analysis.

{During each observation, the two \textit{apec} components have similar best-fit temperatures: a lower-temperature component around 0.8 keV and a higher-temperature component around 2.3 keV. The lower-temperature component is naturally consistent with magnetically confined wind shocks (MCWS) in the secondary's magnetosphere. In the strong-shock approximation, the immediate post-shock temperature is related to the pre-shock velocity through $kT_{\rm s} \approx 1.2 V_8^2$ keV, where $V_8$ is the pre-shock velocity in units of $1000$ km s$^{-1}$ \citep{ud-Doula2014}. The observed temperature of 0.8 keV therefore corresponds to a characteristic shock velocity of approximately 820 km s$^{-1}$. This is compatible with an MCWS origin, although cooler plasma could also arise in the slower-moving wings of the wind-interaction region as the shock wraps around the secondary's magnetosphere. The higher-temperature component, corresponding to a characteristic shock velocity of approximately 1380 km s$^{-1}$, is more likely associated with faster shocks arising from the interaction between the primary wind and the secondary's magnetosphere. In this interpretation, it need not originate from a classical collision between two freely expanding winds. The observed rotational modulation of the X-ray emission provides independent evidence for a substantial magnetospheric contribution, further supporting the MCWS interpretation. Thus, the two-temperature spectrum is consistent with contributions from both magnetically confined and wind-interaction plasma, although the present data do not allow us to uniquely establish the origin of each component.} Although the spectra resemble those of `two-absorber X-ray' sources (e.g. \citealt{Gudel2007}), we did not find evidence for different absorption affecting the two \textit{apec} components when using a `\textit{(phabs*apec)+(phabs*apec)}' model.

\begin{figure}
    \centering
    \includegraphics[width=\linewidth]{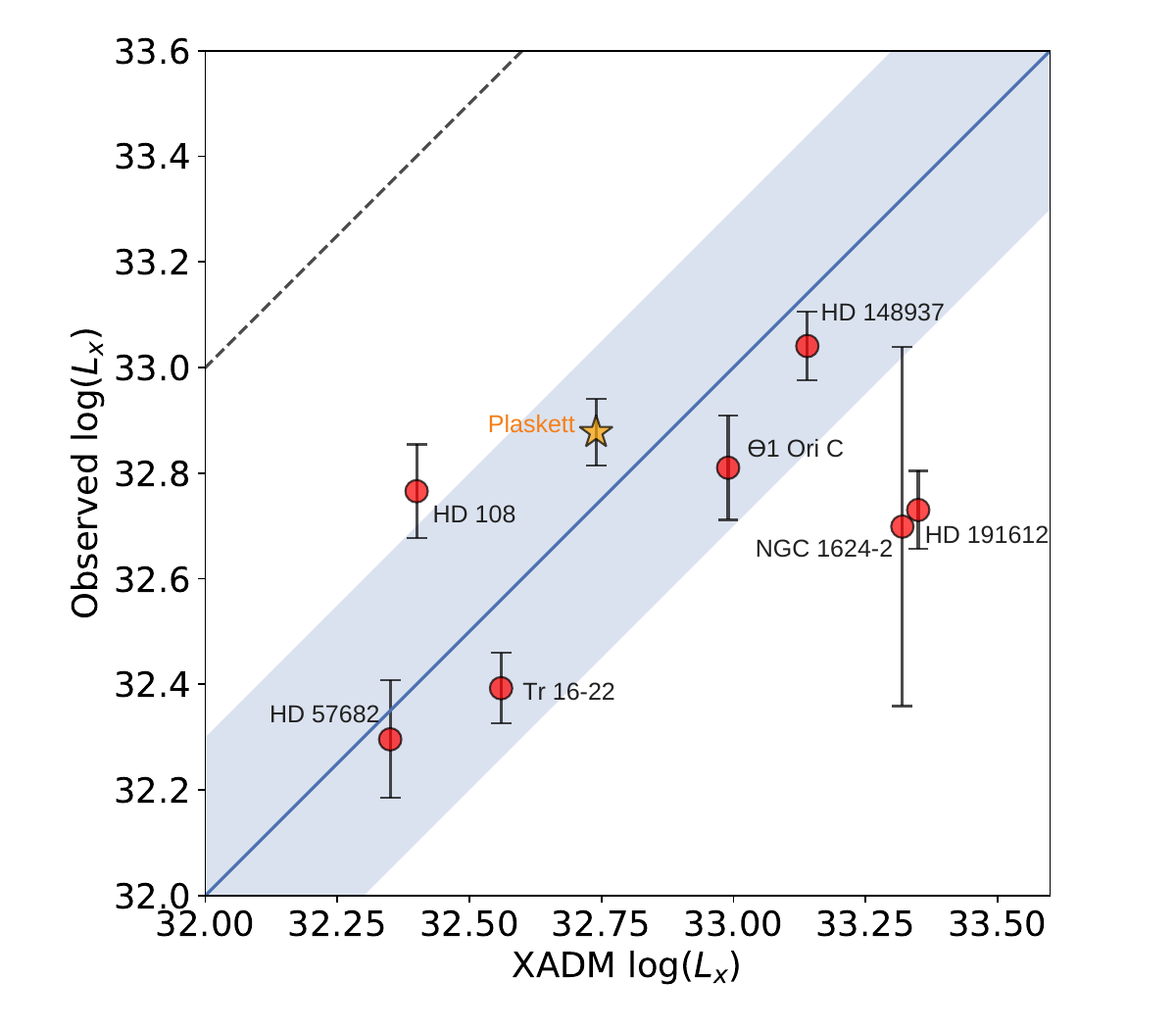}
    \caption{Observed X-ray luminosity of magnetic OB stars vs. the theoretical X-ray luminosity using the XADM model of \cite{ud-Doula2014}. For each star, the luminosity and the corresponding error is re-calculated from \cite{Naze2014} using updated Gaia DR3 distances. The black dotted line represents the ideal XADM model with 100\% efficiency; the light blue shaded area corresponds to an efficiency suggested by magneto hydrodynamic (MHD) models (i.e. 5-20\%). The yellow asterisk represents the position of Plaskett's star in the scaling plot. }
    \label{fig:XADM}
\end{figure}

We also investigated the variation of the model X-ray flux with the orbital phase. The orbital flux modulation is identical to the count rate light curve, with a minimum flux near $\phi_{\rm orb} \sim 0.5$ (Fig. \ref{fig:x_ray_orb2}). We did not notice any significant change in the hardness of the X-ray spectra with the orbital {phase}. The spectra during orbital phases $\sim 0.0$ and $\sim 0.5$ are near-identical in shape, with a similar best-fit temperature, and similar counts near Fe-K lines. The only difference is the overall decrease in flux (or count rate) during phase 0.5 compared to phase 0.0. The normalization for the higher temperature ($N_{\rm kT2}$) remains similar, but we see a significant decrease in the normalization of the cooler component ($N_{\rm kT1}$) during orbital phase 0.5. We did not notice any significant variability of $n_{\rm H}$ with orbital phase, and thus the X-ray orbital variations do not seem associated with a clear absorption change.

\subsection{X-ray Luminosity}

We calculated unabsorbed X-ray luminosity with Gaia DR3 distance of $\approx 1.27$~kpc\footnote{\url{https://vizier.cds.unistra.fr/viz-bin/VizieR?-source=I/352}}. The luminosity varies from $4.5 \times 10^{32}$ to $7.5 \times 10^{32}$ erg s$^{-1}$. The corresponding ratio of X-ray and bolometric luminosity ($L_{\rm bol,1}+ L_{\rm bol,2}$; see Table \ref{tab:parameters}) varies from $\log (L_x/L_{\rm bol}) \sim -6.0$ to $-5.8$. This {X-ray} luminosity is different than the luminosity calculation of \cite{Linder2006} and \cite{Naze2014} from the \xmm data due to the updated Gaia distance. We re-evaluated the position of Plaskett's star in the theoretical scaling relationship for magnetic OB stars, as calculated by \cite{Naze2014}. The prediction is based on the `X-ray Analytical Dynamical Magnetosphere' (XADM) model suggested by \cite{ud-Doula2014}. Fig. \ref{fig:XADM} shows the XADM predictions and the position of known magnetic O stars with detected X-ray emission. It is evident that the luminosity of Plaskett's star is consistent with the 10\% XADM model.

Although the luminosity agrees with the XADM prediction, it should be noted that for Plaskett's star, the $\log (L_x/L_{\rm bol})$ ratio is higher than most O stars (single or binary). From different surveys, it was found that massive star binaries have a slightly larger (0.1--0.2 dex) $L_x/L_{\rm bol}$ ratios \citep{Naze2009, Naze2011}. From a survey of spectroscopically identified galactic O stars, \cite{Nebot2018} determined the peak of the distribution of $\log (L_x/L_{\rm bol})$ around -6.6. {On average, binaries have slightly higher (0.1-0.2 dex) $\log (L_x/L_{\rm bol})$ ratio on average \citep{Rauw2016}, still making Plaskett overluminous.}

\section{High-Resolution X-ray Spectra} \label{sec:xray_spec_high}

High-resolution spectra of Plaskett's star obtained with the \cxo High Energy Transmission Grating Spectrometer (HETGS) were briefly studied by \cite{Pradhan2023}  as part of a survey of high-resolution X-ray spectra from O stars. The authors estimated the line shifts and widths of the major diagnostic lines. For Plaskett's star, the lines are in general found to be unshifted within errors. The error-weighted mean full width half maximum (FWHM) over the Ne X, and Fe XVII lines were found to be consistent with the terminal velocity ($v_{\infty}$) of the star (i.e. $\sim 2450$ km/s). We present an in-depth analysis of the HETGS spectra in the following subsections.

\begin{figure}
    \centering
    \includegraphics[width=\linewidth]{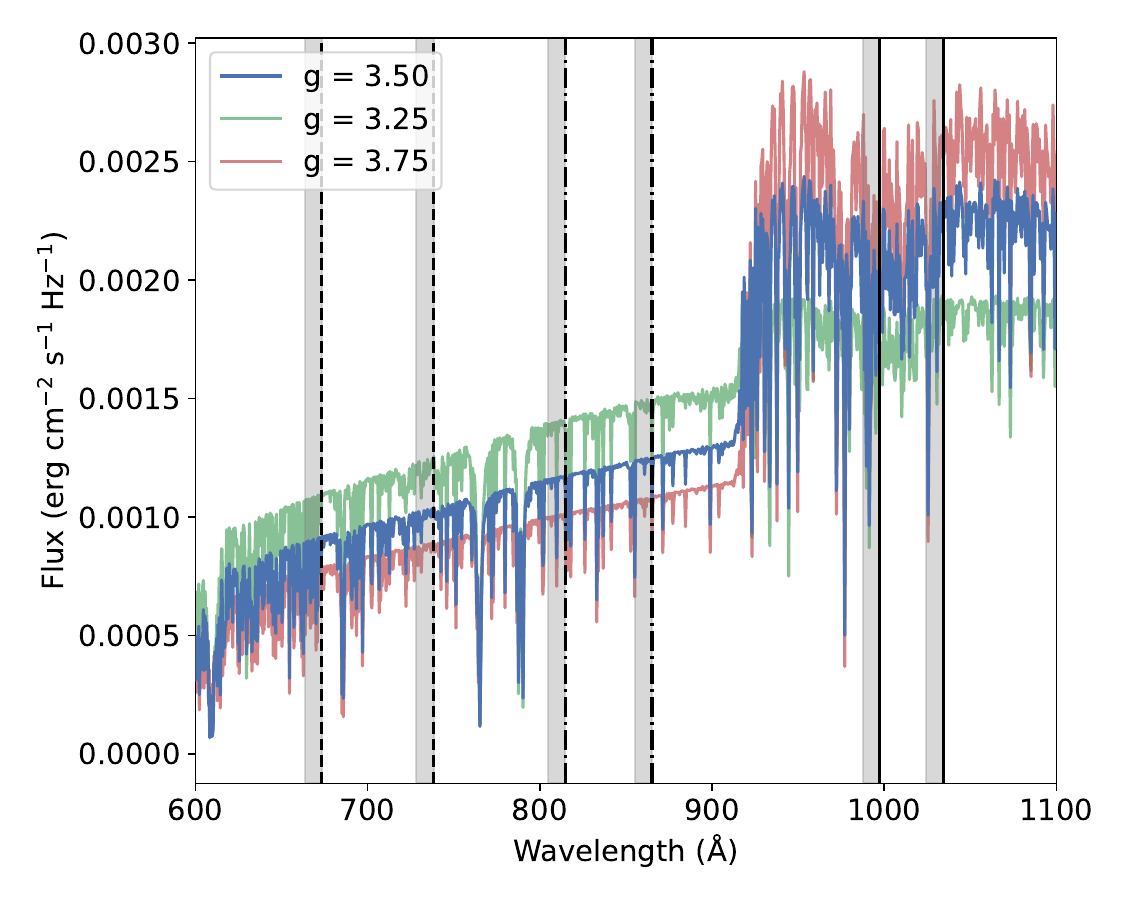}
    \caption{\textsc{TLUSTY}\/ `OSTAR2002' model with $T_{\rm eff} = 32500$, and $\log g = 3.5$ that was used to calculate Eddington flux at the excitation wavelengths shown in blue. The vertical lines represent the driving wavelengths for the UV transitions: the solid lines towards right are for \ion{Mg}{11}, dot-dashed line in the middle for \ion{Si}{13}, and the left-most dotted lines are for S XV. Following the procedure by \cite{Cohen2022} we calculated averaged flux from a range of spectra adjacent to their corresponding wavelengths to account for Doppler shift, and are shaded in gray in the figure. The green and red spectra shows two nearby \textsc{TLUSTY} models with same temperature, but with $\log g = 3.25$ and 3.75, respectively. }
    \label{fig:TLUSTY}
\end{figure}

\subsection{\textit{f/i} Ratio}

The He-like ion complex contains the resonance or $r$ line ($^1{\rm P}_1 \rightarrow ^1{\rm S}_0$), the intercombination or $i$ line ($^3{\rm P}_{2,1} \rightarrow ^1{\rm S}_0$), and the forbidden or $f$ line ($^3{\rm S}_1 \rightarrow ^1{\rm S}_0$).  The transitions $1s^1\, 2s^1\, ^3S_1 \rightarrow 1s^1\, 2p^1\, ^3P_J$ (with J=0,1,2) are in the UV and are dipole transitions. The forbidden-to-intercombination line ratio $\mathcal{R} \equiv f/i$ of He-like ions is diagnostic of the local UV field strength \citep{Gabriel1969,Blumenthal1972}, and thus of the proximity of X-ray-emitting plasma to the surfaces of the two stars \citep{Kahn2001}.  For typical O stars, the undiluted photospheric UV field strength produces a $2s \rightarrow 2p$ transition rate that is much higher than the forbidden decay to ground for ions with $Z \le 10$. The UV photoexcitation rate at the photosphere is typically competitive with the forbidden decay rate for $12 \le Z \le 14$, meaning that the $f/i$ line ratio for \ion{Si}{13} and \ion{Mg}{11} is sensitive to plasma locations within a few stellar radii of the photosphere.

\begin{figure}
\centering
\gridline{\fig{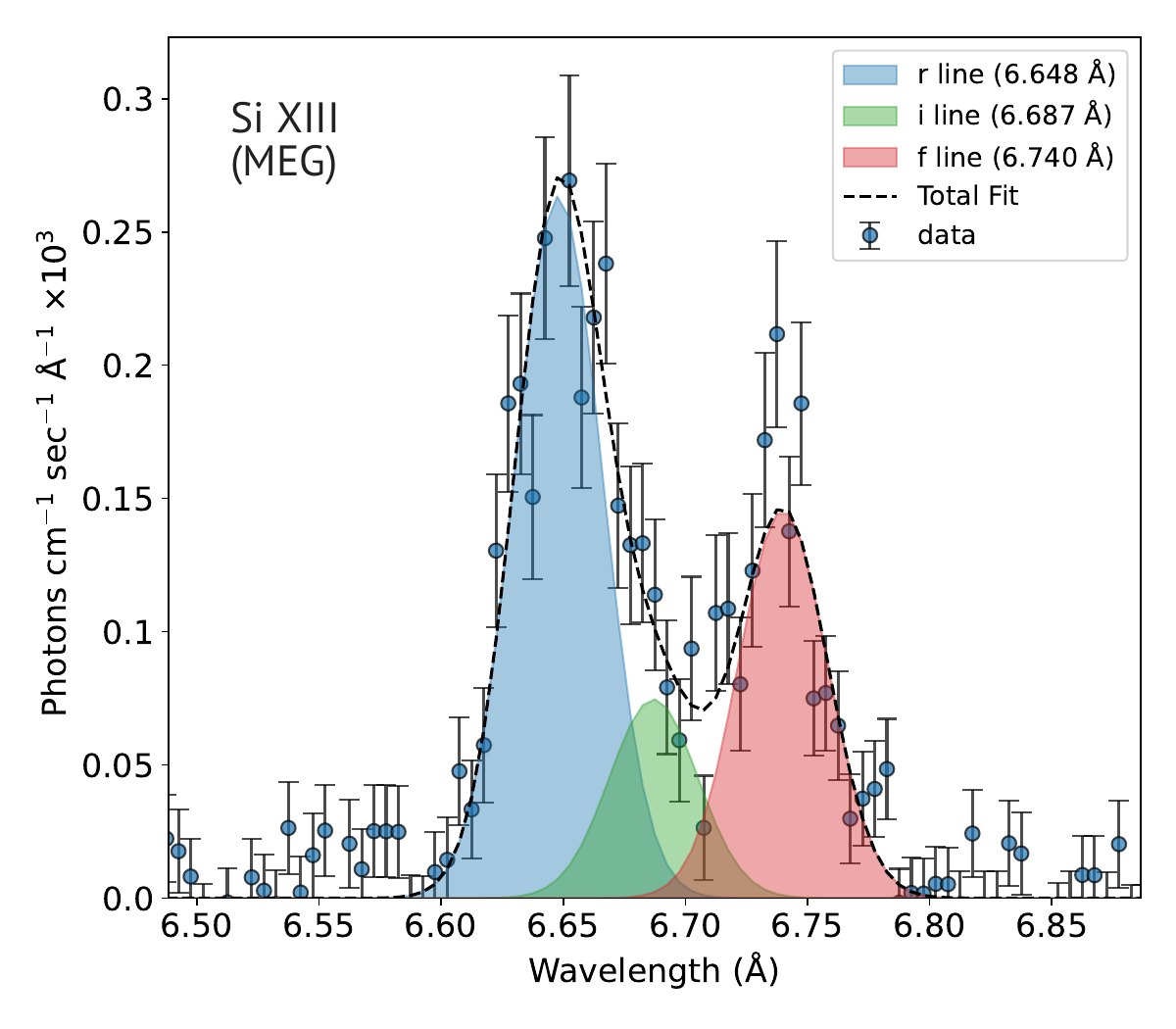}{0.375\textwidth}{(a)} \label{fig:MEG_Si}}
\gridline{\fig{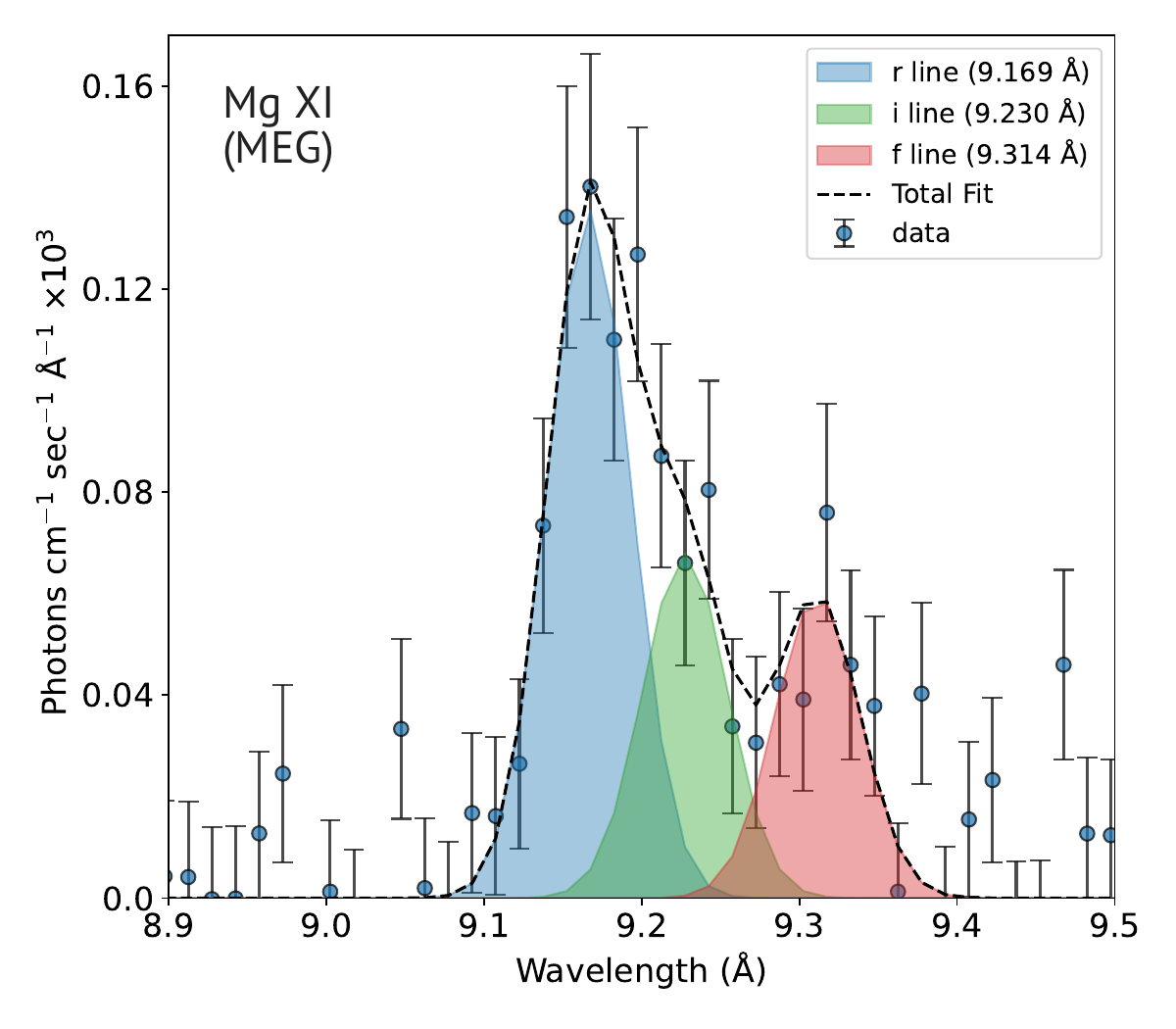}{0.375\textwidth}{(a)} \label{fig:MEG}}
\gridline{\fig{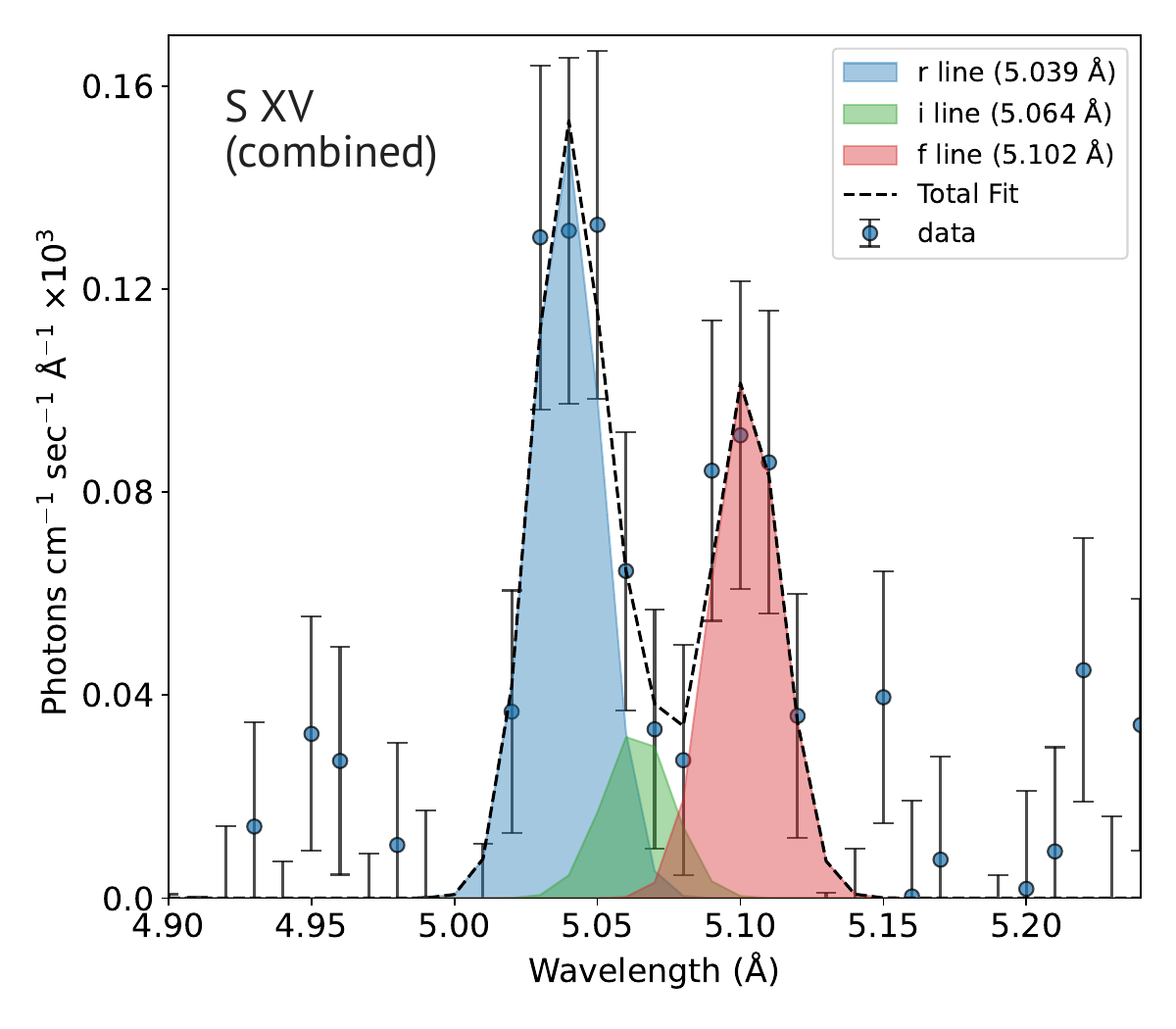}{0.375\textwidth}{(b)} \label{fig:HEG}}
    \caption{Baseline subtracted MEG and HEG spectra of the \ion{Si}{13} and \ion{Mg}{11} triplet. The blue, green and red gaussians represent the best fit to the recombination ($r$), inter-combination ($i$), and forbidden ($f$) line, respectively. The black dashed curve represents the combined model fit to the spectrum.   }
    \label{fig:SiXIII_both}
\end{figure}

\subsubsection[Radial Dependence of R(r)]{Radial Dependence of $\mathcal{R}(r)$}

The dependence of the $f/i$ ratio, i.e., $\mathcal{R}(r)$ on atomic parameters and UV mean intensity can be described using equations \citep{Blumenthal1972, Leutenegger2006}:

\begin{equation} \label{fi_1}
    \mathcal{R}(r) =  \frac{R_0}{1 + n_e(r)/N_c + 2P W(r)}, 
\end{equation}

\noindent
where $P = \phi_*/\phi_c$, $\phi_*$ being the photo-excitation rate of $2 \ ^3S_1 \rightarrow 2 \ ^3P_J$, given by:

\begin{equation}  \label{fi_2}
    \phi_* = 8 \pi \frac{\pi e^2}{m_e c} f \frac{H_{\nu}}{h \nu}, 
\end{equation} 

\noindent
where $\phi_c$ is the critical photoexcitation rate, $H_{\nu}$ is the surface Eddington flux at the excitation wavelength, $f$ is the sum of oscillator strengths of the given transitions, $\nu$ is the UV photon's frequency.  In equation \ref{fi_1}, $n_e(r)$ is the X-ray electron density, $N_c$ is the critical density, and $R_0$ represents the low density and $P=0$ limit. For O stars, the radiation term is dominant except for very close to the photosphere. Thus, equation \ref{fi_1} can be simplified as \citep{Leutenegger2006}:

\begin{equation} \label{fi_3}
    \mathcal{R}(r) =  \frac{R_0}{1 + 2P W(r)}, 
\end{equation}

\noindent
where $W(r)$ is the radial-dependent geometric dilution factor, given by:

\begin{equation} \label{fi_4}
    W(r) = \frac{1}{2} \left(  1 - [1-(R_*/r)^2]^{1/2}  \right)
\end{equation}

The values of $R_0$ were taken from \cite{Cohen2022}, and the values for $f$, $\phi_c$, and $\nu$ for relevant transitions are taken from \cite{Leutenegger2006}. In order to get UV flux at the driving wavelength for relevant transitions, we used {\textsc{TLUSTY}} model atmospheres. We adopted a model similar to the parameters of Plaskett's components ($T_{\rm eff} = 32500$ K, $\log g = 3.5$). The UV flux obtained from this method has several limitations (e.g. see \citealt{Cohen2022}), and the adopted values are approximations only. In Fig. \ref{fig:TLUSTY}, we show the {\textsc{TLUSTY}} model used for this work, together with its nearest models. To obtain $H_{\nu}$, we averaged the {\textsc{TLUSTY}} model flux over a wavelength range to account for the Doppler broadening (shown in Fig. \ref{fig:TLUSTY}). The most sensitive line complex for these studies with \cxo HEG and MEG spectra are the S XV, \ion{Si}{13}, and \ion{Mg}{11} lines near 5.0, 6.7, and 9.2 Å. The average $H_{\nu}$ values for the two transitions of these given ions are -- S XV: 0.00100, 0.00079; \ion{Si}{13}: 0.00123, 0.00114; \ion{Mg}{11}: 0.00204, 0.00184 (erg cm$^{-1}$ s$^{-1}$ Hz $^{-1}$).

\begin{figure}
\centering
\gridline{\fig{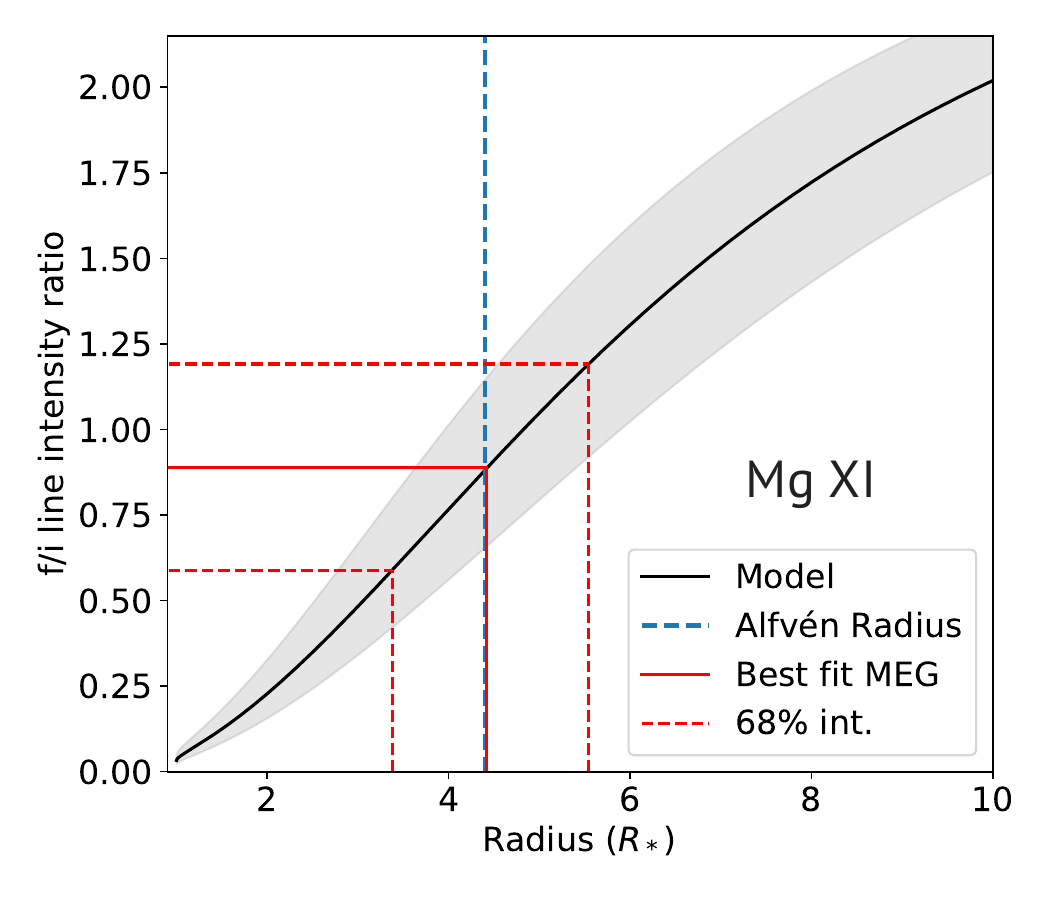}{0.95\linewidth}{(a)} \label{fig:fi1}}
\gridline{\fig{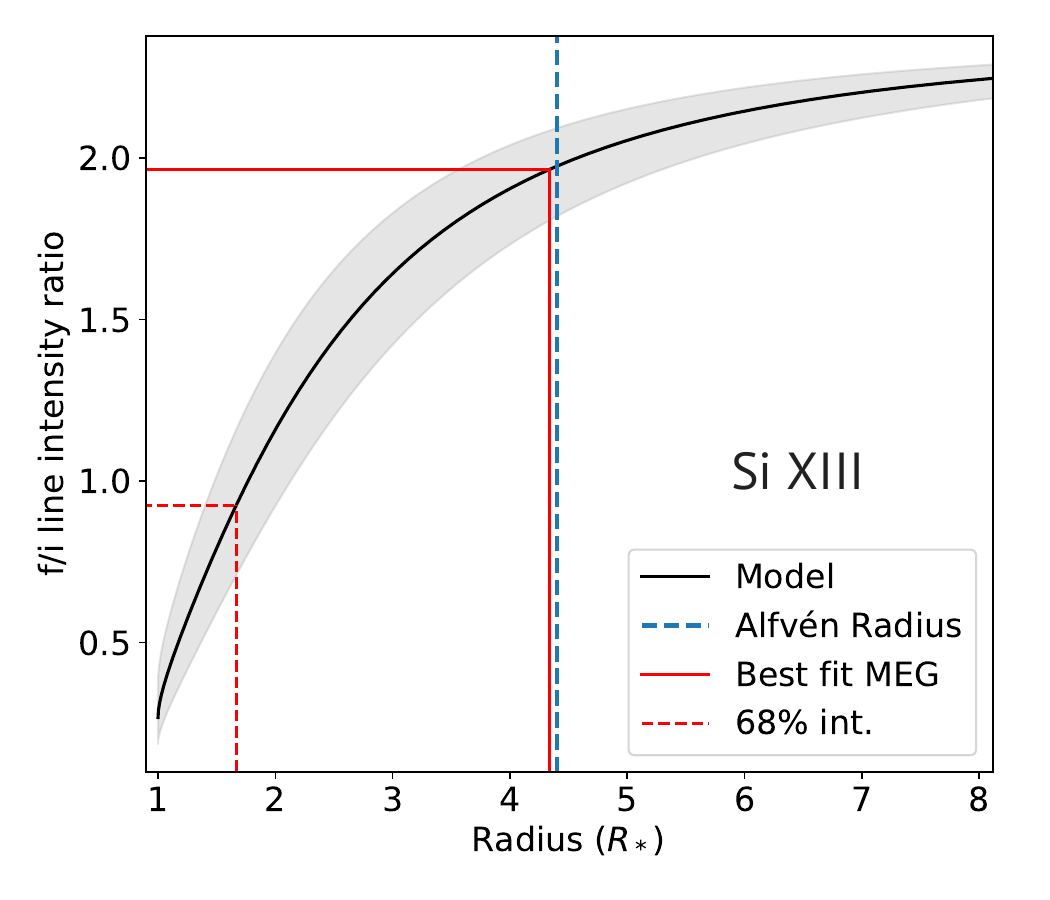}{0.95\linewidth}{(b)} \label{fig:fi2}}
\caption{The radius of X-ray onset, or formation radius $R_{f/i}$ vs. $f/i$ ratio. The black curve is the model X-ray onset radius { (in units of the secondary's stellar radius)}, calculated using equations \ref{fi_1}--\ref{fi_4}, depending on the $f/i$ ratio ($\mathcal{R}(r)$). Panel (a) and (b) shows the model for \ion{Mg}{11} and \ion{Si}{13} line complexes, respectively. The grey dashed line are calculated using photospheric flux of $0.67 \ H_\nu$ and $1.5 \ H_\nu$, assuming it to be the uncertainty in the model. The solid red line in each case represent the best-fit $f/i$ ratio obtained from spectral fitting. The dashed lines similarly represents the 68\% confidence intervals. For \ion{Si}{13}, the upper limit is not shown as it exceeds $R_0$ (see eqn. \ref{fi_3}). The blue vertical dashed line shows the Alfv{\'{e}}n radius in each case.   \label{fig:model_fi}}
\end{figure}

In our case, the S XV line was too weak to get a reliable line ratio measurement, while the \ion{Si}{13} and \ion{Mg}{11} lines show noticeable emission from all three components (i.e., $r$, $i$, and $f$). We fit data from two grating arrays (HEG and MEG) combining two spectral orders ($-1$ and $+1$). The lines were fit after subtracting a baseline flux, and the fitting was performed using MCMC analysis. Details about the MCMC analysis can be found in Appendix \ref{App:mcmc}. {In all our analysis, contribution from satellite lines are ignored. We also note that Ne~{\sc x} lines can contribute within the Mg~{\sc xi} triplet wavelength range; however, given their lower emissivity, we do not explicitly include this contribution in our line fitting.} Sample spectra of the \ion{Si}{13}, \ion{Mg}{11}, and S XV lines with a three-component Gaussian fit are shown in Fig. \ref{fig:SiXIII_both}. For \ion{Mg}{11}, we obtained $\mathcal{R} = 0.89_{-0.30}^{+0.31}$, and for \ion{Si}{13}, we obtained $\mathcal{R} = 1.96_{-1.04}^{+1.02}$. In both cases, we get the ratio $(f+i)/r = \mathcal{G}$ to be $\sim 1$ for both \ion{Si}{13} and \ion{Mg}{11} lines. This ratio corresponds to a hot plasma with temperature $\sim 10^7$ K \citep{Porquet2000}.

\subsubsection{Location of the X-ray Emitting Plasma}

\begin{figure*}
\centering
\gridline{\fig{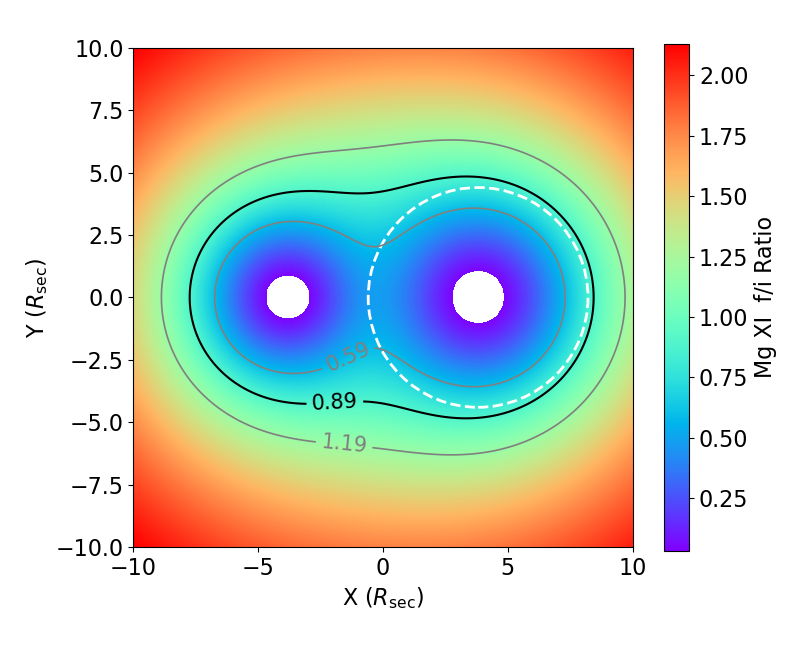}{0.492\linewidth}{(a) \ion{Mg}{11} (combined)} \label{fig:2D_map1}
          \fig{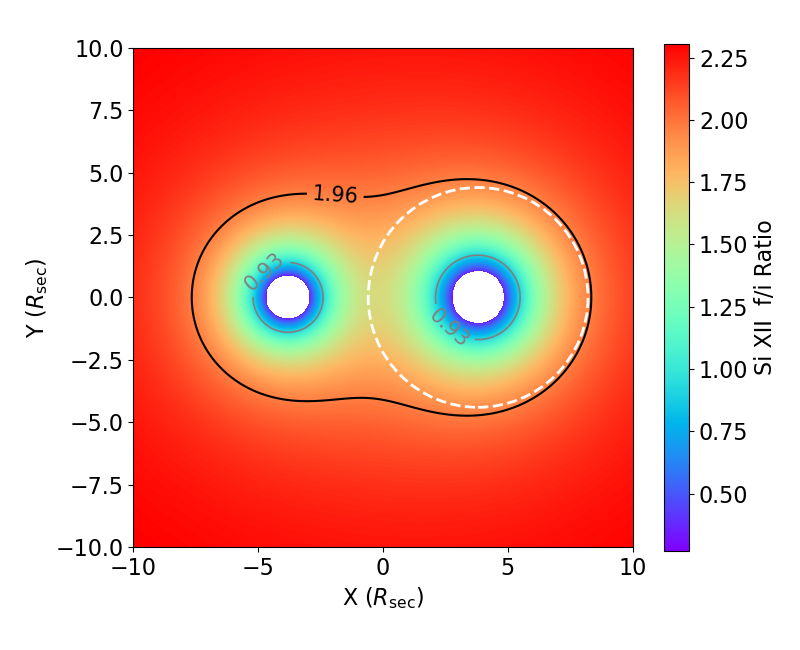}{0.492\linewidth}{(b) \ion{Si}{13} (combined)} \label{fig:2D_map2}}
    \caption{Map of $f/i$ ratio along the orbital surface of the Plaskett's star system calculated from (a) \ion{Mg}{11} and (b) \ion{Si}{13} lines. In both cases, the primary is on the left. The separations and radii are approximately to scale. The white dotted circles represent the Alfv{\'{e}}n surface of the magnetic secondary. The black solid curves represent the best-fit value of $f/i$ ratio from the corresponding He-like triplets. The gray curves represent the $68\%$ confidence intervals from MCMC fit (if available). }
    \label{fig:2D_map}
\end{figure*}

Before calculating the region of X-ray onset using equations \ref{fi_3} and \ref{fi_4}, we ensured that the approximation of $n_e (r) << N_c$ is correct. For an O star with magnetosphere, the density can reach $\sim 10^{12} \ {\rm cm}^{-3}$ \citep{ud-Doula2014}. Even in that case, $n_e (r)/N_c < 0.1$, having negligible impact on the measured $f/i$ ratios. At first, we consider a scenario in which X-ray emission is generated only from the magnetic secondary. If we consider a simplified spherically symmetric density profile, using the procedure described in the previous section, we obtain the radial variation of $f/i$ ratio given in Fig. \ref{fig:model_fi}. The black curves in Fig. \ref{fig:model_fi} represent the radial variation of the corresponding lines, while the gray bands represent the model uncertainty. The model uncertainties include the model atmosphere uncertainties in UV flux scaled by 0.67 (lower limit) and 1.5 (upper limit). The uncertainty calculated in this manner is only representative and possibly overestimated. For both \ion{Mg}{11} and \ion{Si}{13} line, the onset radius from the best-fit $f/i$ value is fully compatible with the Alfv{\'{e}}n radius of the magnetic star (shown with the blue dashed lines in Fig. \ref{fig:model_fi}). 

Instead of a 1-D variation, we also map the variation of the $f/i$ ratio including the contribution of the primary component (with similar UV contribution) along the orbital surface. The 2-D variation of $f/i$ ratios along the orbital plane is shown in Fig. \ref{fig:2D_map}. The lower limits of the line ratios are compatible with emission below the Alfv{\'{e}}n surface, and the best fit line shows the emission onset site near the Alfv{\'{e}}n surface. However, although there is hardly any chance of getting emission at the line-of-center, given the uncertainty, {we cannot exclude the possibility that the X-rays may also have a contribution from the interaction between the primary wind and the magnetosphere of the secondary}. In addition, because of radiative braking, it is possible that the colliding wind region does not emit along the line of center, but along the wings.

In the MCWS scenario of X-ray production from single massive magnetic stars, MHD simulations suggest the primary origin of X-rays just below the Alfv{\'{e}}n surface, where the magnetically confined plasma is co-rotating with highest kinetic energy \cite{ud-Doula2014}.  In case of the Plaskett star, the uncertainty in both the Alfv{\'{e}}n and the line-ratio measurement remains high, and thus a definitive location cannot be assumed. Additionally, winds are expected to have a non-uniform density. In case of Plaskett's secondary, maximum X-ray producing clouds are expected to form near the junction of magnetic and rotational equator. Thus, a complex, realistic model of X-ray production is required. Furthermore, the $f/i$ ratio modeling does not take into account the X-rays produced as a result of the colliding-wind shocks.

\begin{figure*}
\centering
\gridline{\fig{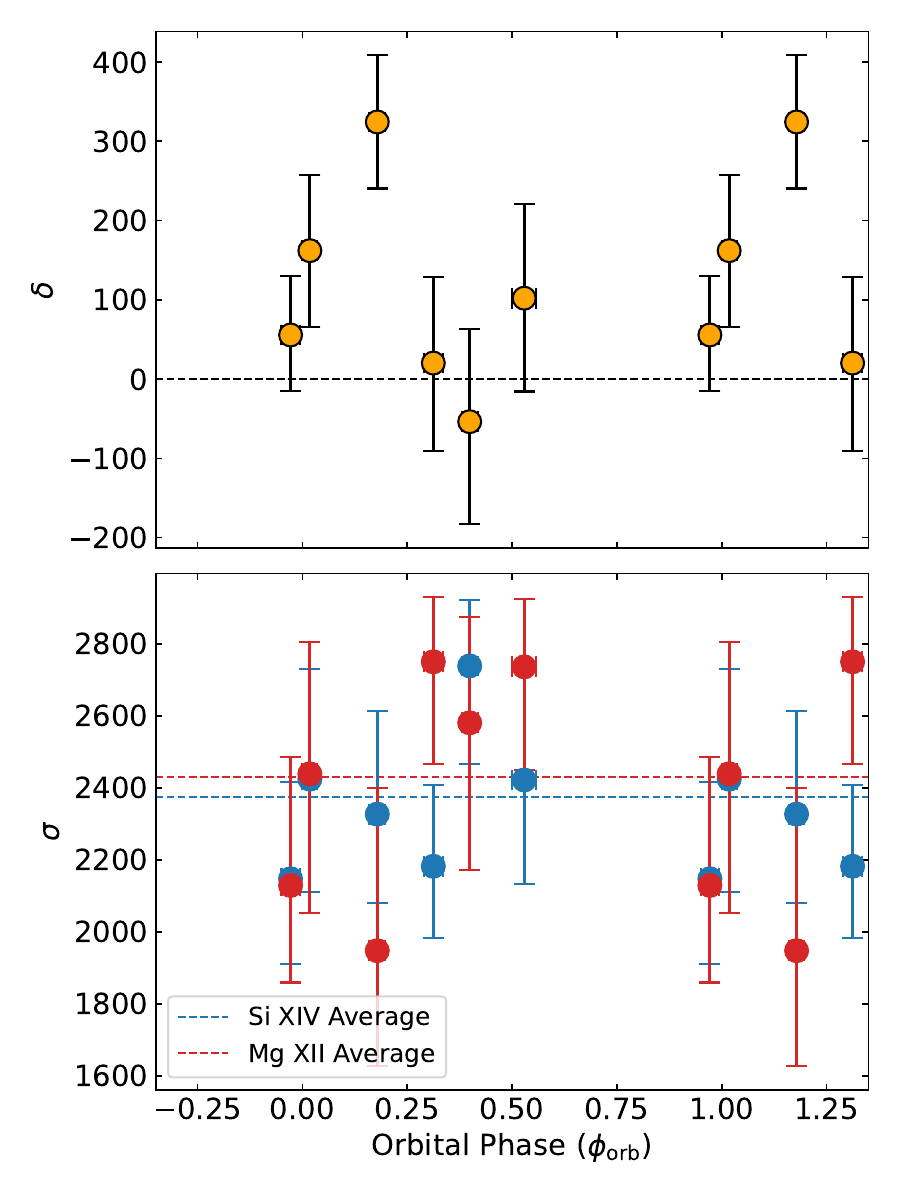}{0.42\textwidth}{(a) Orbital Variation ($P_{\rm orb} \sim 14.4$ d)} \label{fig:ophase_velocities}
          \fig{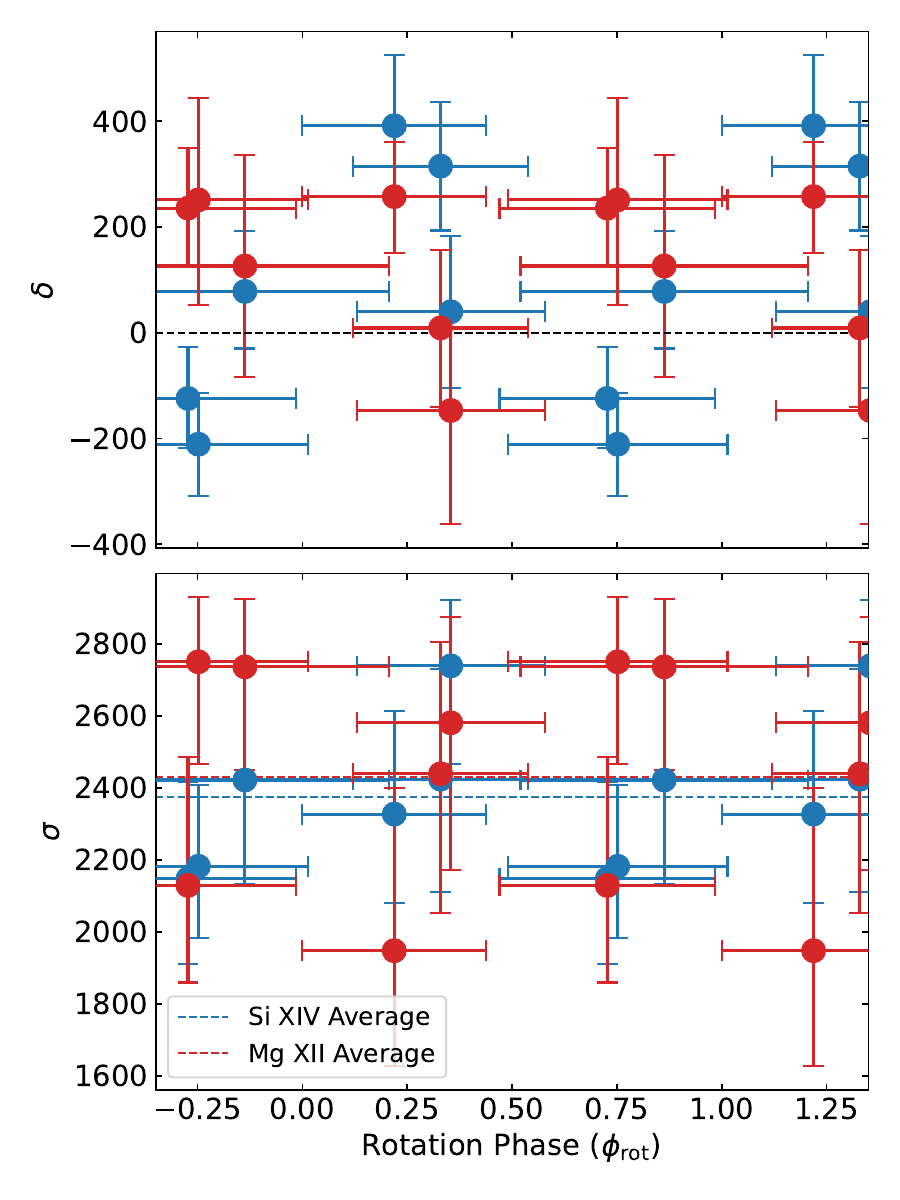}{0.42\textwidth}{(b) Rotational Variation ($P_{\rm rot} \sim 1.22$ d)} \label{fig:rphase_velocities}}
    \caption{(a) Best fit Gaussian width $\sigma$ and centroid shift $\delta$ for the Ly$\alpha$ lines of \ion{Si}{14} and \ion{Mg}{12} as a function of orbital phase. (b) Same as (a), but as a function of rotational phase of the broad lined star.  }
    \label{fig:phase_velocities}
\end{figure*}

\subsection{Line Profile Variability}

In Fig. \ref{fig:phase_velocities}, we show the results of fitting Gaussians to the two strongest unblended lines (\ion{Si}{14} and \ion{Mg}{12}) in the \cxo spectra of Plaskett's star. { The average line width ($\sim2400\,{\rm km\,s^{-1}}$) is comparable to, although somewhat larger than, the primary wind terminal velocity of $v_{\infty,1}=2300\pm100\,{\rm km\,s^{-1}}$ inferred by \cite{Wade2026} for the stripped narrow-lined primary, based on its UV P-Cygni profiles. The line width therefore does not require an additional velocity component substantially exceeding the characteristic wind velocity of the system.} We do not observe any statistically significant variation in line width and line shift with either orbital or rotational phase. If we assume the magnetic secondary to be the main X-ray emitter, the centroid shifts with orbital phase are expected to be equivalent to the RV variation obtained from optical measurements. \cite{Wade2026} found that the RV variation from the magnetic component has an amplitude of the order of $10 \ {\rm km \ s}^{-1}$, which is much smaller than the measurement uncertainty of our data. It is impossible to measure such an orbital variation of centroid shifts with current instrumental resolution and sensitivity.

For the magnetic secondary, if we consider the Rigidly Rotating Magnetosphere (RRM; \citealt{Townsend2005}) model, two co-rotating plasma clouds are expected to form at the intersections between the rotational and magnetic {equatorial planes}, where they are at magnetohydrostatic equilibrium. If X-rays are produced strictly in these clouds, we could have seen a faint variation of line shifts comparable to the equatorial velocity of the star along the rotation phase. A variation of the circumstellar magnetosphere that stretches $\pm500$ km/s with rotational phase was seen from H$\alpha$ emission \citep{Grunhut2022}. However, we do not see such variation, partly because of the poor signal to noise, and possibly because the distribution of X-ray producing plasma is not concentrated simply on these two clumps but rather has a complex structure. Such complex structures were previously suggested by \cite{Carciofi2013} where they implemented a 'dumbbell+disk' model of the magnetospheric plasma distribution to explain the polarimetric signature in the magnetic massive star $\sigma$ Ori E.

\section{Radio} \label{sec:radio}

\subsection{Broad-band Radio Spectra}

\cite{Kurapati2017} previously detected radio emission from this star in the Ka (32 GHz, or 1 cm) and X (10 GHz, or 3 cm) bands at orbital phases 0.14 and 0.47, respectively. These observations correspond to 0.51 and 0.80 rotation phases. Because the observations were carried out at different epochs and phases (orbital or rotational), an unambiguous determination of the spectral index is not possible. The authors reported an intraband spectral index of $-1.2 \pm 1.0$ between the sub-bands of the X band, and $0.0 \pm 2.5$ from the Ka band. { The mass-loss rate inferred by interpreting the VLA emission entirely as thermal free-free emission is substantially higher than several theoretical estimates, although the magnitude of the discrepancy depends on the adopted wind prescription.}

\begin{figure}
    \centering
    \includegraphics[width=\linewidth]{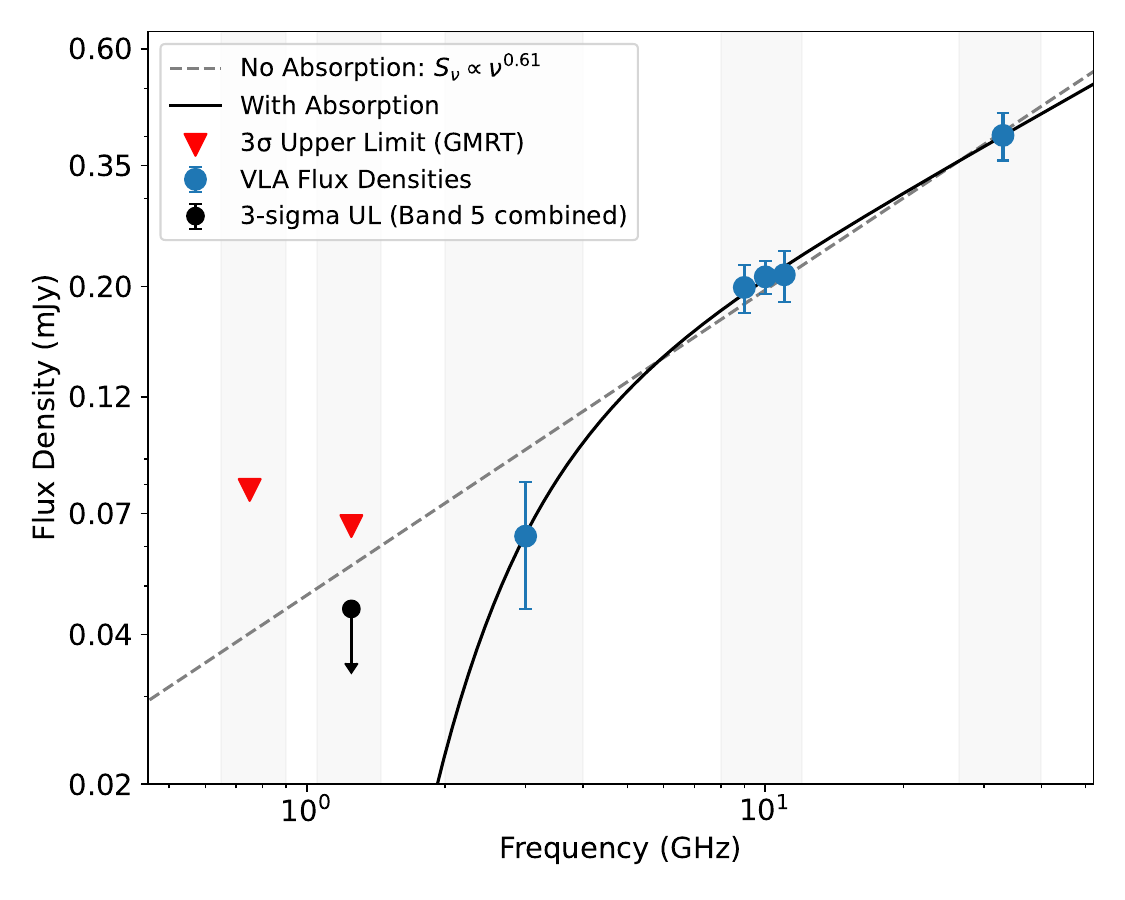}
    \caption{Broad band radio spectra of Plaskett's star. The blue circles with error bars represent re-calculated VLA flux densities, while the red triangles represent the typical $3-\sigma$ upper limits from uGMRT observation. The black point with downward arrow represents the deeper upper limit from uGMRT Band 5 data obtained from combing all datasets. The black dashed line gives a simple power-law with spectral index $+0.61$. The black solid curve represents the model flux density after including absorption. The gray vertical bands in the background show the position of uGMRT/VLA frequency bands.}
    \label{fig:radio_spectra}
\end{figure}

Our reanalysis of the VLA data revealed additional detection in the S band with a flux density of $0.063 \pm 0.018$ mJy (Fig. \ref{fig:radio_spectra}). We divided the X-band into several sub-bands to recalculate the intraband spectral index. We found the flux densities of $0.199 \pm 0.022$ mJy, $0.209 \pm 0.016$ mJy, and $0.211 \pm 0.025$ mJy in the sub-bands centered at 9, 10, and 11 GHz, respectively. These values resulted in a spectral index of $\alpha_{\rm intra} = 0.3 \pm 0.8$. { Because of its large uncertainty this value is consistent with the inter-band spectral index} of $\alpha_{\rm inter} = 0.61 \pm 0.06$. Although the target was detected in all VLA observations, it remained undetected at uGMRT frequencies at all observing epochs. For individual observations, we reached an image rms noise level of $\sim30 \ \mu$Jy in individual observations. After combining all observations, we obtained a composite image with noise level of $\sim 15 \ \mu$Jy, where the target was again undetected. We also searched archival radio observations and surveys (e.g. RACS survey, Table \ref{tab:radio_obs}), but did not find any detection. The radio spectrum of the Plaskett's star is shown in Fig. \ref{fig:radio_spectra}.

\subsection{Emission Mechanism and Non-detections}

\begin{figure}
    \centering
    \includegraphics[width=0.47\textwidth]{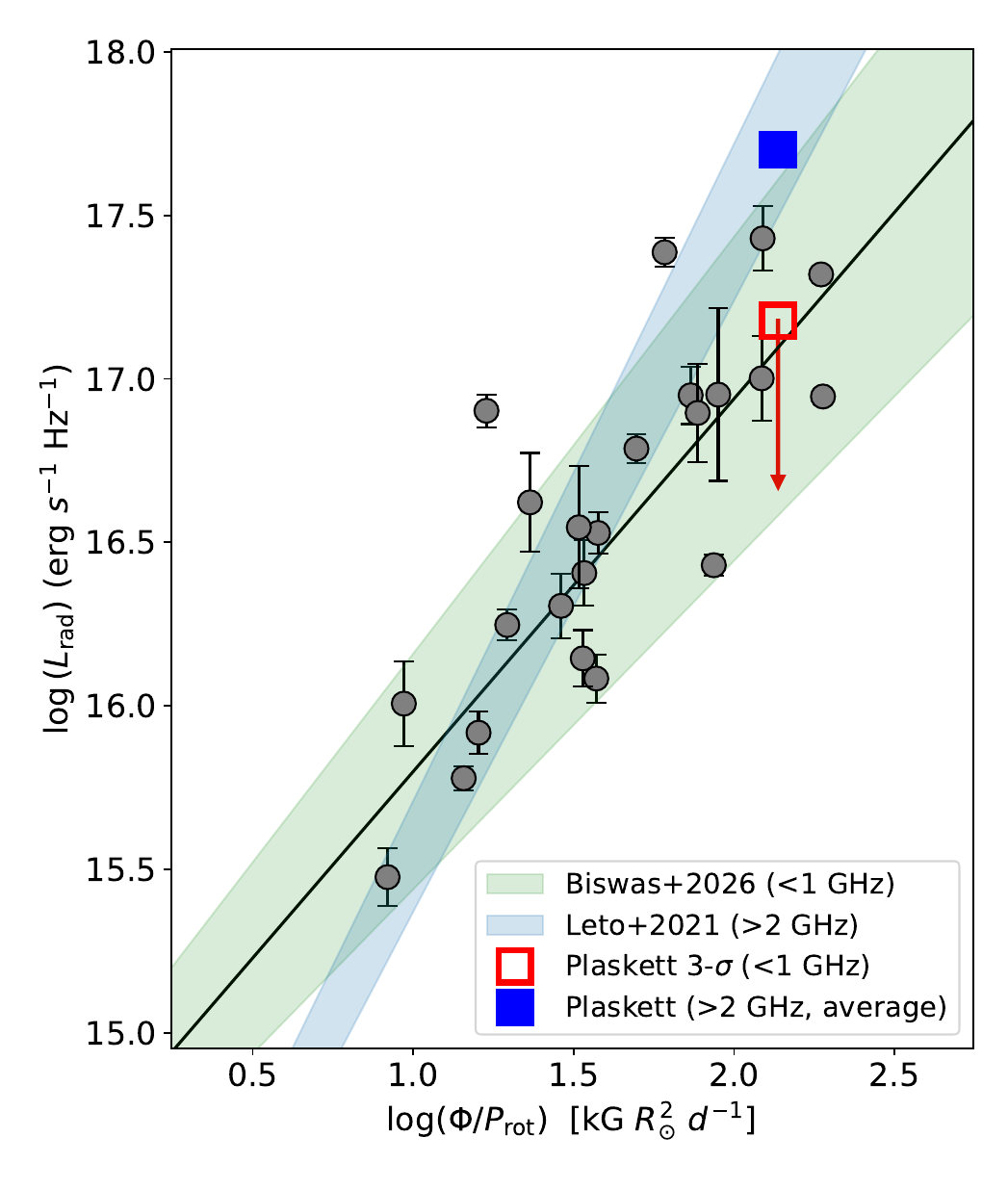}
    \caption{Location of Plaskett's star in the scaling relationship among other magnetic massive stars (gray circles) observed at uGMRT Band 4 \citep{Biswas2026}. The x-axis shows $\log(\Phi / P_{\rm rot})$ in units of kG~$R_{\odot}^2$~${\rm d}^{-1}$, where $\Phi$ is the surface magnetic flux and $P_{\rm rot}$ is the stellar rotation period. The black line and the light green band represent the best fit scaling relation, while the blue band represents the scaling trend observed from high frequency ($>2$ GHz) studies by \citep{Leto2021}. The average luminosity from high frequency VLA observations of Plaskett is shown in blue square, where the low frequency non-detection with 3-$\sigma$ upper limit is shown in red hollow square and downward arrow. }
    \label{fig:radio_scaling}
\end{figure}

At first, we investigated the contribution of free-free absorption in the flux density using the equation \citep{Bieging1989}: 

\begin{equation}
    S_{\nu, \rm thermal} = 23.2 \  (\gamma \  g_{\rm ff} \nu )^{2/3} D^{-2} \left( {\dot{M} \ Z}/{v_{\infty}} \right)^{4/3} \ {\rm Jy},
\end{equation}

\noindent
where $S_{\nu, \rm thermal}$ is the free-free flux density in Jy, $g_{\rm ff}$ is the free-free gaunt factor, $Z$ is the mean ionic charge, {$\gamma$ is the mean number of electrons per ion}, $\mu$ is the mean ionic weight and $D$ is the distance from the star in units of kpc. We used the theoretical mass-loss rates { discussed in Sec. \ref{OrbMod}} and found $S_{\nu, \rm thermal}$ of the order of a few $\mu$Jy in all bands. This value is insignificant with respect to the observed flux in the VLA bands. Conversely, if the emission is assumed to be free-free emission, we get predicted mass-loss rates to be {at least one, and possibly two orders of magnitude higher} than the theoretical mass-loss rate, making this assumption nonphysical. Although the intra-band spectral index is consistent with both thermal and nonthermal interpretations given its uncertainty, the observed radio luminosity, together with the magnetic-field/rotation scaling relation and the expected strong free-free absorption at low frequencies, supports a substantial nonthermal contribution. Nonthermal radio emission from this star can either come from the magnetosphere of the secondary via the gyrosynchrotron mechanism or from synchrotron emission from the collision of wind of the two components.

In recent surveys of radio emission from magnetic massive stars, it is found that the gyrosynchrotron emission scales with magnetic field strength, stellar radius, and rotation period (\citealt{Leto2021, Shultz2022, Biswas2026}). We investigated the position of Plaskett's star in the scaling relationship among other magnetic massive stars observed at uGMRT Band 4 \citep{Biswas2026}. In Fig. \ref{fig:radio_scaling}, we show the position of our target with respect to scaling laws at both low and high frequencies. It is evident that the observed luminosity of Plaskett at high frequencies is consistent with the scaling relationship derived at high frequencies by \cite{Leto2021}. The upper limit of radio luminosity at the sub-GHz frequency is also near the expected luminosity of the system based on the low-frequency scaling law. A possible reason for this non-detection  at sub-GHz frequency is free-free absorption, where photons are absorbed in the presence of an ion. The high mass-loss rates of O stars along with free-free absorption can make the wind optically thick, making a `radio photosphere' \citep{Wright1975}. For massive stars, we can define the radius of this free-free radio photosphere (free-free radius, $R_{\rm ff}$) as the radius where the free-free optical depth is unity. For a non-magnetic spherically symmetric wind, we can estimate the radius following \cite{Torres2011}:

\begin{equation}
    R_{\rm ff} = 70.71 \left( \frac{\dot{M}_{-8}}{ \nu_{GHz}  v_{\infty, 8} } \right)^{2/3}  f^{1/3} T^{1/2}_5, 
\end{equation}

\noindent
where $R_{\rm ff}$ is the free-free radius in units of $3 \times 10^{12}$ cm, $\dot{M}_{-8}$ is the mass-loss rate (in units of $10^{-8} M_{\odot}/{\rm yr}$), $\nu_{\rm GHz}$ is the frequency in GHz, and $f$ is the clump volume filling factor, and $T_5$ is the temperature in units of $10^5$ K. The expected value of $R_{\rm ff}$ is higher than the Alfv{\'{e}}n radii ($R_{\rm A}$) in all frequency bands of our study, especially at low frequencies. In general, $R_{\rm ff} >> R_{\rm A}$ implies that any magnetospheric emission will be absorbed within the radio photosphere of the star. At sub-GHz wavelengths, $R_{\rm ff}$ is also larger than orbital separation ($R_{\rm ff}/R_{\rm A} \ \sim 42$, at Band 4), which suggests that any non-thermal emission from colliding winds is also getting absorbed at low frequencies, thus explaining the non-detection. In addition to gyrosynchrotron emission, coherent radio emission is also expected from stars hosting a centrifugal magnetosphere via electron cyclotron maser emission (ECME; \citealt{Trigilio2000, Das2022, Das2022c}). In case of Plaskett's star, the dense wind is suspected to absorb most of this emission. According to the scaling relation of \cite{Das2022c}, we expect an ECME flux of few $\mu$Jy from the secondary of Plaskett's star, which is below the detection threshold of our observations.

To account for free-free absorption, we introduced an additional attenuation term: $ S_{\nu} = S_0 \left({\nu}/{\nu_0}\right)^{\alpha} e^{-\tau_{\nu}}, $ where the optical depth $\tau_{\nu}$ is modeled as $\tau_{\nu} = \tau_0 \left({\nu}/{\nu_0}\right)^{-2.1}$. Here, $\tau_0$ represents the optical depth at $\nu_0$, and the exponent $-2.1$ corresponds to the frequency dependence of the free-free opacity in a plasma. We performed non-linear least squares fitting to determine the best fit values of $\alpha$, $S_0$, $\tau_0$, and $\nu_0$. The power-law fit alone resulted in a spectral index of $\alpha \approx 0.61$, while incorporating free-free absorption yielded a modified spectral index $+0.52$ and an optical depth of $\tau_0 \approx 2.14$ at $\nu_0 = 1.64$ GHz. The latter model provided a better fit, particularly at lower frequencies, where deviations from a pure power law are observed (Fig. \ref{fig:radio_spectra}).

\section{Conclusion} \label{sec:Discussion_Conclusion}

In this work, we investigated the X-ray and radio properties of the magnetic massive binary Plaskett's star with the goal of separating magnetospheric emission from other possible mechanisms. We combined phase-resolved \textit{Chandra} ACIS and HETGS observations with a reanalysis of archival VLA data, as well as new uGMRT observations. This multi-wavelength approach provides new constraints on the relative roles of the magnetic secondary and the wind interaction zone, and clarifies the origin of the observed high-energy emission. The main findings of this study are summarized below.

\begin{enumerate}

    \item We confirm that the orbital phase-folded X-ray light curve shows a decline near phase 0.5. This phase corresponds to a configuration in which the primary star and its wind lie in front of the magnetic secondary. The observed drop in count rate is therefore consistent with partial occultation by wind absorption of the dominant X-ray source, in agreement with previous X-ray observations. However, since we did not notice any significant variability of $n_{\rm H}$ with orbital phase, the orbital variations of the X-rays do not seem to be associated with a clear absorption change.

    \item The rotational phase-folded X-ray light curve shows significant variability that is independent of orbital modulation. The light curve morphology broadly resembles that of optical magnetospheric diagnostics, indicating that the X-ray emission responds to changes in the viewing geometry of the centrifugal magnetosphere. Rotational X-ray modulation is commonly observed among magnetic O-type stars and is generally interpreted within the framework of oblique magnetic rotator models. 

    \item We confirm that the broad-band X-ray spectrum is characterized by at least two thermal plasma components. Fits to the ACIS spectra show that a single-temperature model cannot reproduce the observed spectral shape, whereas a model including two optically thin thermal components provides a satisfactory description of the data. The lower-temperature component, at approximately 0.8~keV, is compatible with expectations for magnetically confined wind shocks associated with the magnetic secondary and contributes the majority of the observed X-ray flux. The higher-temperature component, near 2.5~keV, may additionally include contributions from plasma heated in the wind--wind collision region of the binary system.

    \item Modeling of the He-like triplets in the HETGS spectra suggests that at least part of the X-ray emitting plasma is located relatively close to the magnetic secondary, at radii comparable to the characteristic Alfv\'{e}n scale of the magnetosphere. However, the uncertainties on the inferred formation radii remain substantial, and the current measurements do not uniquely distinguish between magnetospheric and colliding-wind contributions. 

    \item Gaussian fits to the unblended \ion{Si}{14} and \ion{Mg}{12} lines in the \cxo\ spectra show an average line width of $\sim 2400\ \mathrm{km\ s^{-1}}$, which is lower than the value expected for the stars in a non-magnetic scenario. Within the uncertainties, we find no statistically significant variation in either line width or centroid shift with orbital or rotational phase. The expected radial velocity variations are too small to be resolved with the current data, and the present measurements therefore do not place strong constraints on the detailed spatial distribution of the X-ray emitting plasma. The absence of rotational variation of line profiles can be attributed partly to the poor signal to noise ratio, and possibly because the distribution of X-ray producing plasma has a complex structure.

    \item Reanalysis of VLA data produces detections across multiple bands and spectral analysis confirms the presence of nonthermal radio emission. However, with the available phase coverage, we cannot yet distinguish between magnetospheric and colliding-wind contributions. Given that $R_{\rm ff}/R_{\rm A}$ is smaller at these wavelengths, colliding winds have a better chance of being detected in these wavelength ranges.

    \item The system remains undetected at all uGMRT frequencies, even after combining all observations. The predicted free-free photosphere at sub-GHz frequencies is significantly larger than both the Alfv\'{e}n radius and the orbital separation. Therefore, any nonthermal radiation at these frequencies, whether produced in the magnetosphere or in the colliding-wind region, is expected to be fully absorbed by the dense O-star winds, which explains the non-detections. The dense wind is also expected to absorb all coherent emission produced via ECME from the centrifugal magnetosphere of the magnetic secondary.

\end{enumerate}

Together, the X-ray and radio diagnostics presented here provide a more refined picture of Plaskett's star and the relative roles of magnetospheric confinement and binary interaction. The combination of orbital and rotational X-ray variability, the soft dominant plasma component, and the He-like triplet formation radii all point to the magnetosphere of the secondary star as the primary source of the system's X-ray emission. The detection of nonthermal radio emission confirms the presence of energetic electrons, although the strong free-free opacity of the dense O-star winds explains the absence of any low-frequency counterpart in the uGMRT bands. {However, a deeper, near-simultaneous multi-band observations and multi-epoch monitoring are needed to better constrain the radio spectrum and establish the presence of nonthermal emission, either through the detection of polarization or rotational radio variability.} Continued phase-resolved X-ray and radio monitoring, combined with future three-dimensional MHD modeling, will be essential for a more complete description of the magnetospheric plasma and wind distribution in a massive, interacting binary such as Plaskett's star.


\begin{acknowledgments}
A.B. acknowledges support from the Department of Atomic Energy, under project "Next Generation Instrumentation for Radio Astronomy" (project UID number RTI4017). A.B. acknowledges support from Dr. Colin Folsom who helped generate TLUSTY models for this work. G.A.W. acknowledges support in the form of a Discovery Grant from the Natural Sciences and Engineering Research Council (NSERC) of Canada. Y.N. acknowledge support from the FNRS and the university of Liège. This research has made use of data obtained from the Chandra Data Archive provided by the Chandra X-ray Center (CXC). The GMRT is run by the National Centre for Radio Astrophysics of the Tata Institute of Fundamental Research. The authors thank GMRT staff for helping with the observations. The GMRT/uGMRT data used in this work is available from the GMRT online archive (GOA, \url{https://naps.ncra.tifr.res.in/goa/data/search}). The National Radio Astronomy Observatory and Green Bank Observatory are facilities of the U.S. National Science Foundation operated under cooperative agreement by Associated Universities, Inc.  \textit{TESS} data is readily available from the Mikulski Archive for Space Telescopes (MAST, \url{https://archive.stsci.edu/}), doi: \dataset[10.17909/s10t-4d57]{https://doi.org/10.17909/s10t-4d57}.
\end{acknowledgments}

\vspace{5mm}
\facilities{XMM, Chandra, GMRT, JVLA, TESS }

\software{CASA \citep{McMullin2007},
          Astropy \citep{astropy:2013, astropy:2018, astropy:2022}  
          numpy \citep{harris2020array}, 
          Period04 \citep{Lenz2005},
          scipy \citep{Virtanen2020},
          CARTA \citep{Carta2021},
          CIAO \citep{Fruscione2006},
          SAS \citep{Gabriel2004},
          XSPEC \citep{Arnaud1996}
          }


\bibliography{sample631}{}
\bibliographystyle{aasjournal}


\appendix

\begin{figure*}
\centering
\gridline{\fig{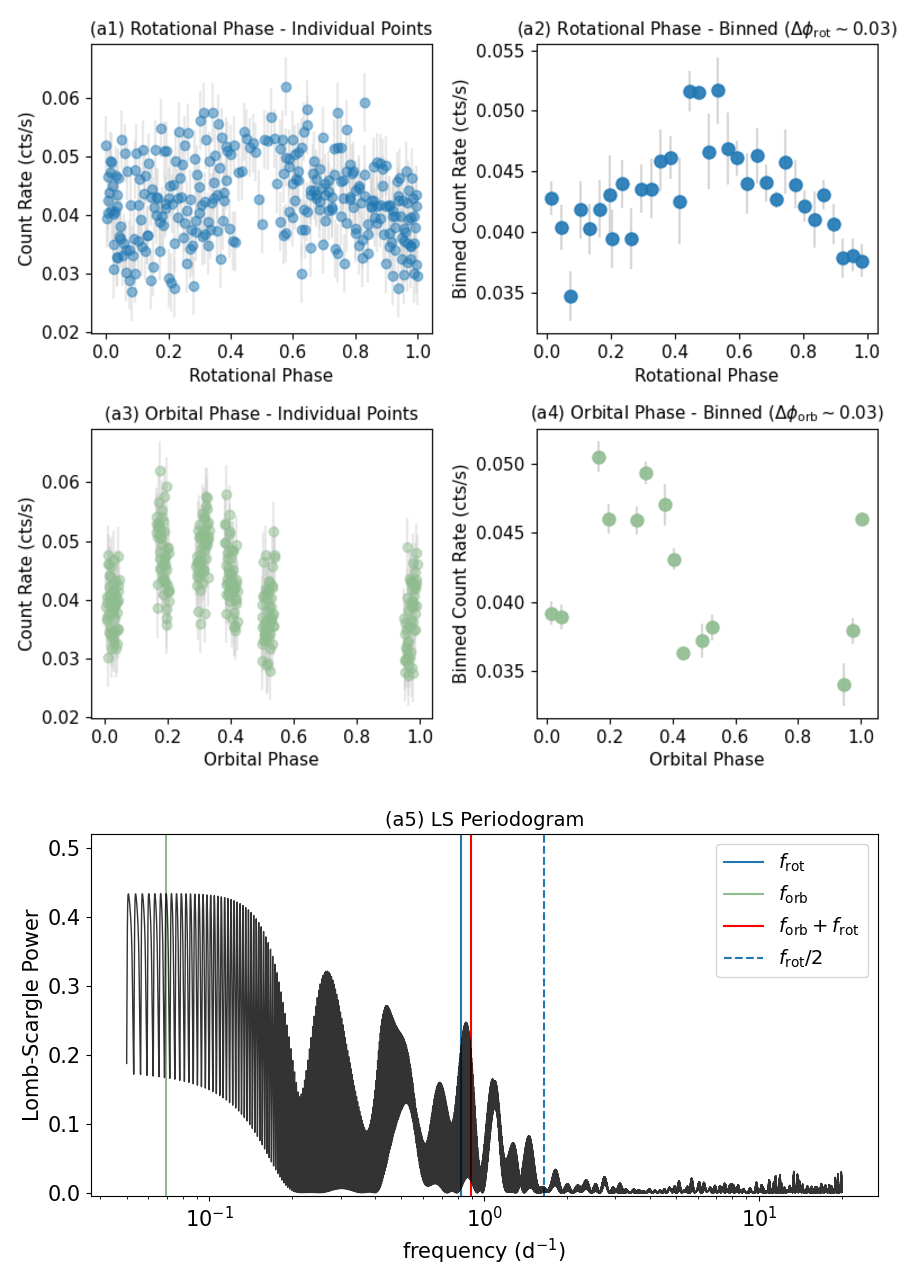}{0.46\textwidth}{(a) Synthetic: With signal.} \label{fig:simulated_signal}
          \fig{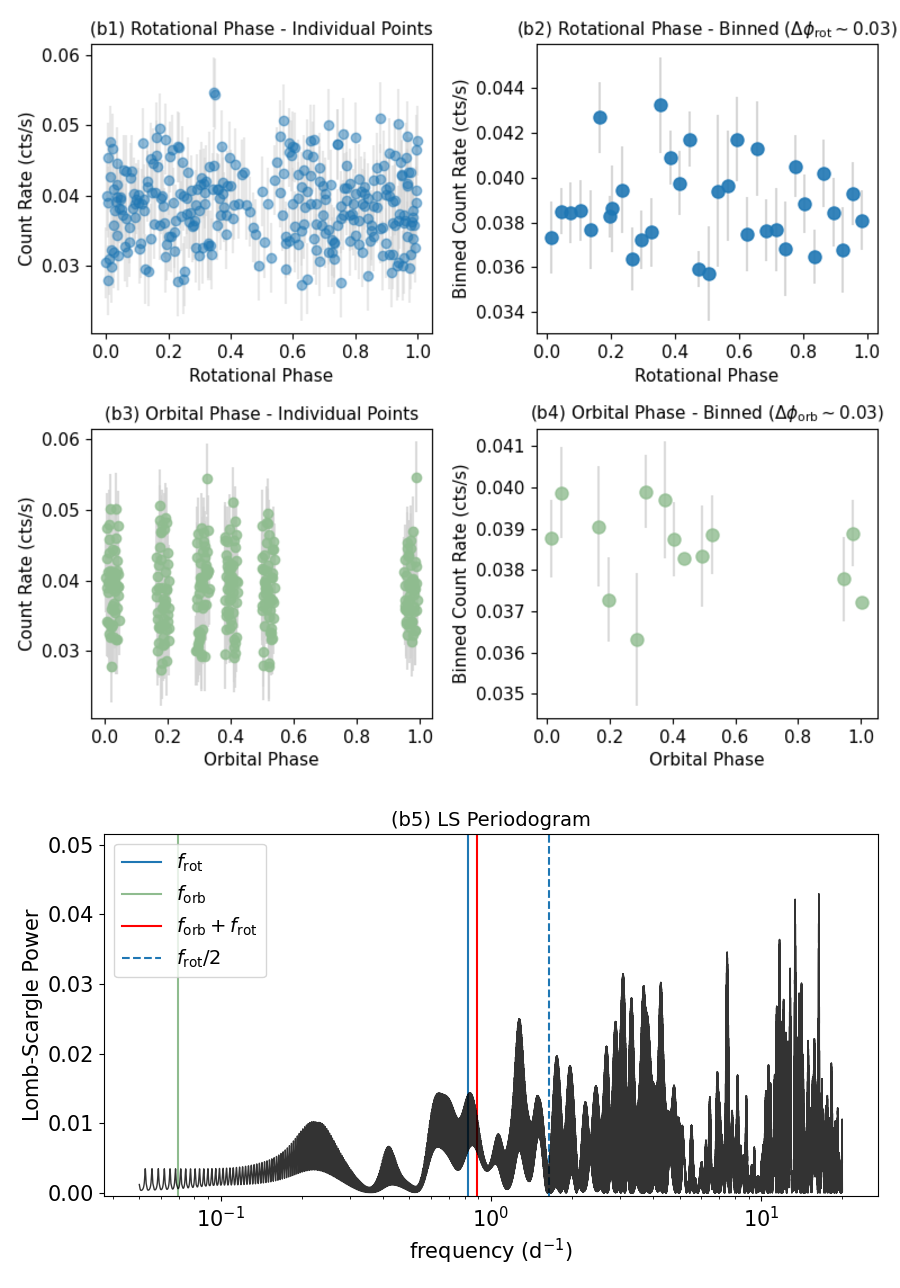}{0.46\textwidth}{(b) Synthetic: Only noise.} \label{fig:simulated_noise}}
\gridline{\fig{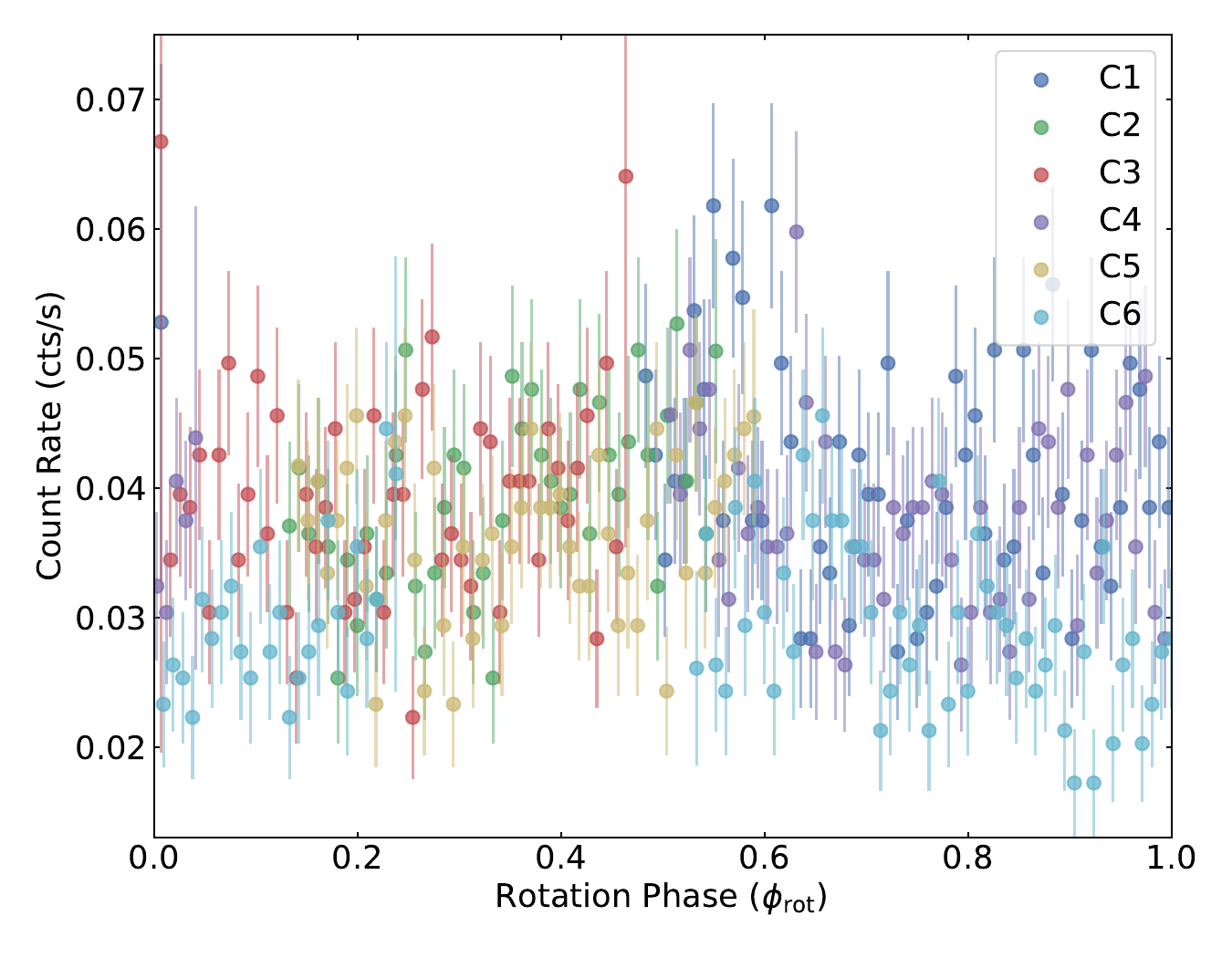}{0.4\textwidth}{(c) Observed individual 1000 sec light curves} \label{fig:x_ray_rot_all}
\fig{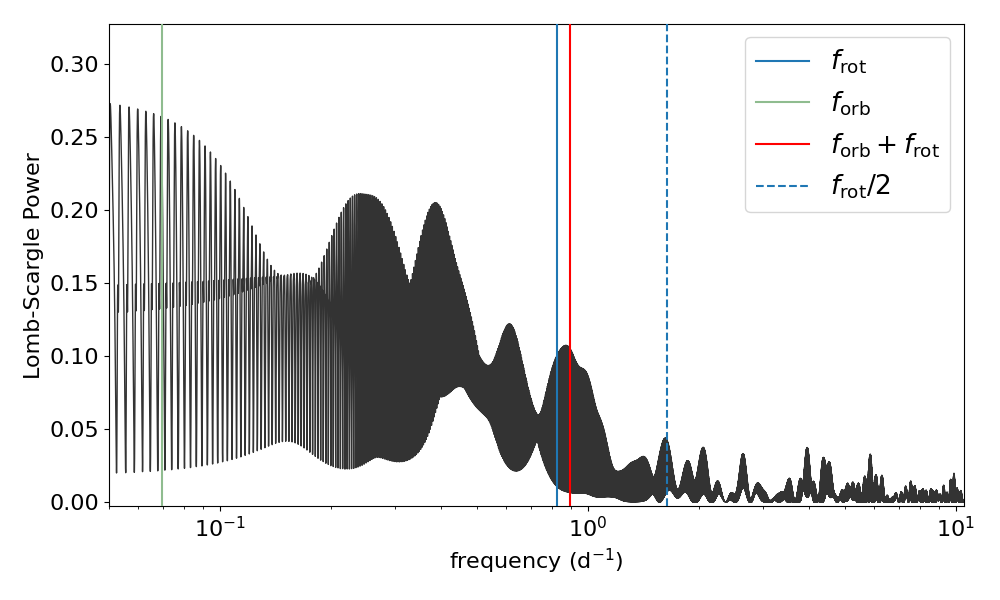}{0.54\textwidth}{(d) Observed periodogram.} \label{fig:observed_signal}}
    \caption{\textbf{(a--b)} Synthetic light curve with exact same HJDs, period and noise level with (a) intrinsic rotational+orbital signal, and (b) just noise. This figure emphasizes the necessity of a combined binning of the light curve to extract the orbital and rotational signal separately. In both (a) and (b), the top left and middle left panels show the rotational phase folded 1 ks light-curve without binning multiple observations together. In the 2nd and 4th panel of each case, they are binned in phase combing multiple observations, showing a clear detection of the injected variation in (a) and non-detection of any clean variation in (b). In the bottom panels, the Lomb-Scargle periodograms are shown. repetition of the same procedure shows a peak between $f_{\rm rot}$ (blue solid line) and  $f_{\rm rot} + f_{\rm rot}$ (red solid line) for case (a). \textbf{(c-d)} The observed light curve and periodogram for comparison. }
    \label{fig:simulated}
\end{figure*}

\section{Synthetic Light Curves to Study Binning Effects} \label{App:synthetic}

In order to demonstrate the need for binning multiple observations in order to detect rotational modulation, we produce synthetic light curves with (a) both rotational and orbital signal along with noise, and (b) only noise on top of a constant emission. For the first case, we assume simple sinusoidal variation (instead of a double-wave pattern for the rotation) with appropriate variability amplitude. We inject a random noise level similar to the observed case. Phased light curves with respect to the rotational (a1, b1) and orbital (a3, b3) periods are shown in Fig. \ref{fig:simulated}. The binned and phased light curves combining multiple observations are also shown (a2, a4, b2, b4) along with their Lomb-Scargle (LS) periodograms (a5, b5). Although the variation was not evident in the 1 ks light curves, it was evident after binning multiple observations in the first case. The injected signal is also marginally detected in the first LS periodogram between $f_{\rm rot}$ and $f_{\rm rot} + f_{\rm rot}$. This trend is detected consistently in all repetitions. The observed periodogram also shows a very similar trend.

\section{Additional Details from Low-Resolution Spectral Modeling} \label{App:spectra}

\begin{figure*}
\centering
\gridline{\fig{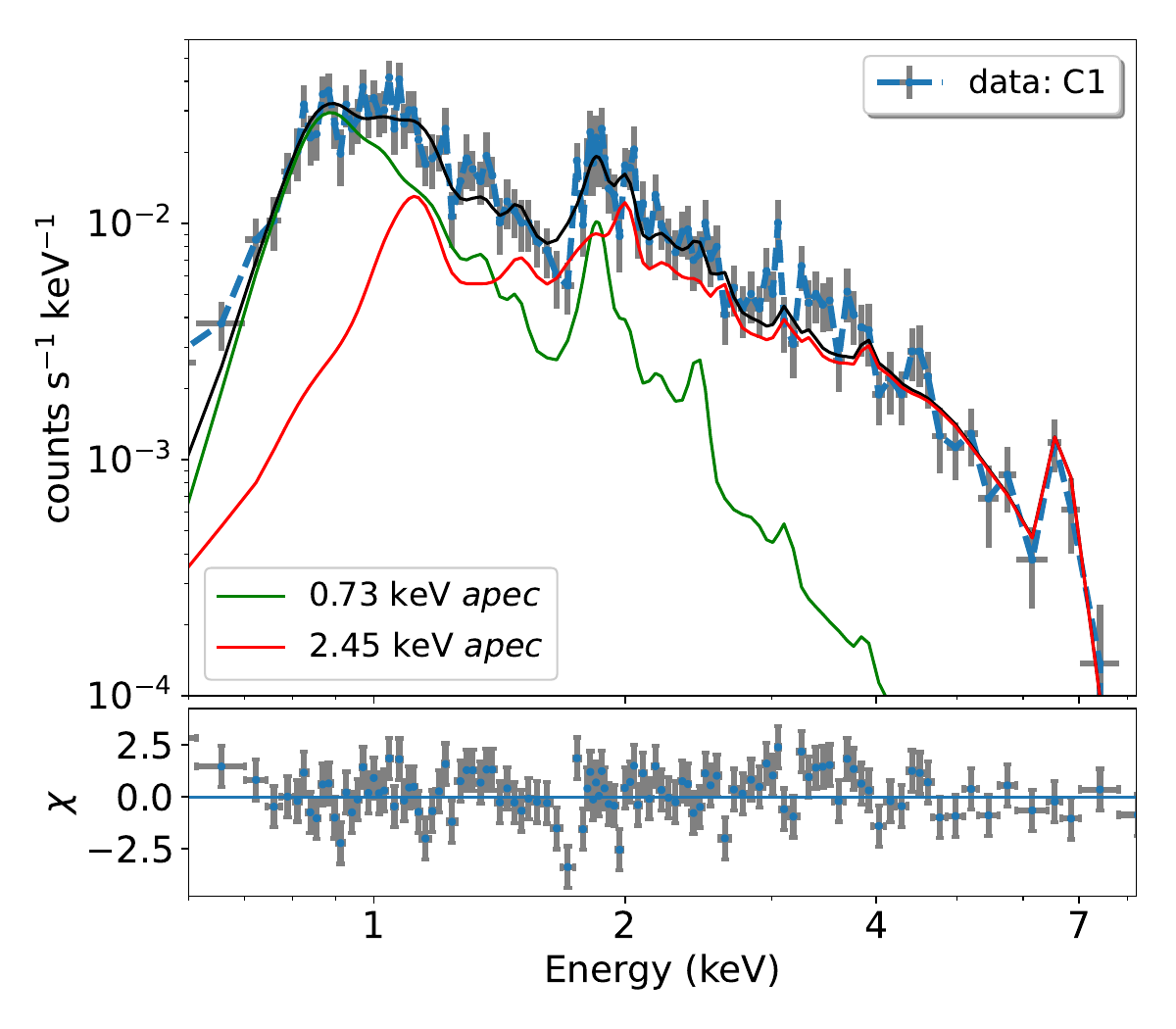}{0.33\textwidth}{(a) Data ID: C1, $\phi_{orb}= 0.95$} \label{fig:s1} 
          \fig{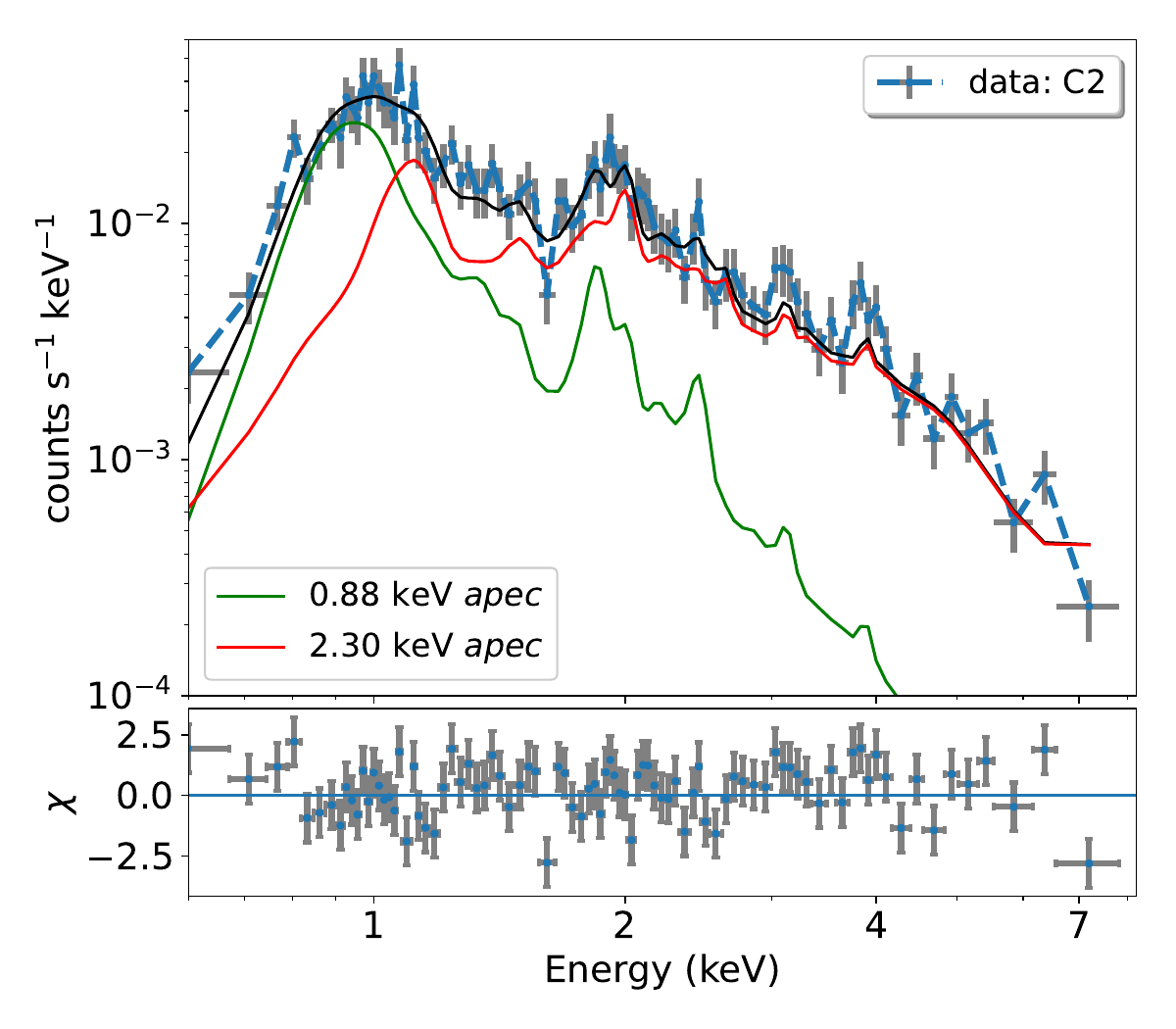}{0.33\textwidth}{(b) Data ID: C2, $\phi_{orb}= 0.00$} \label{fig:s2} 
          \fig{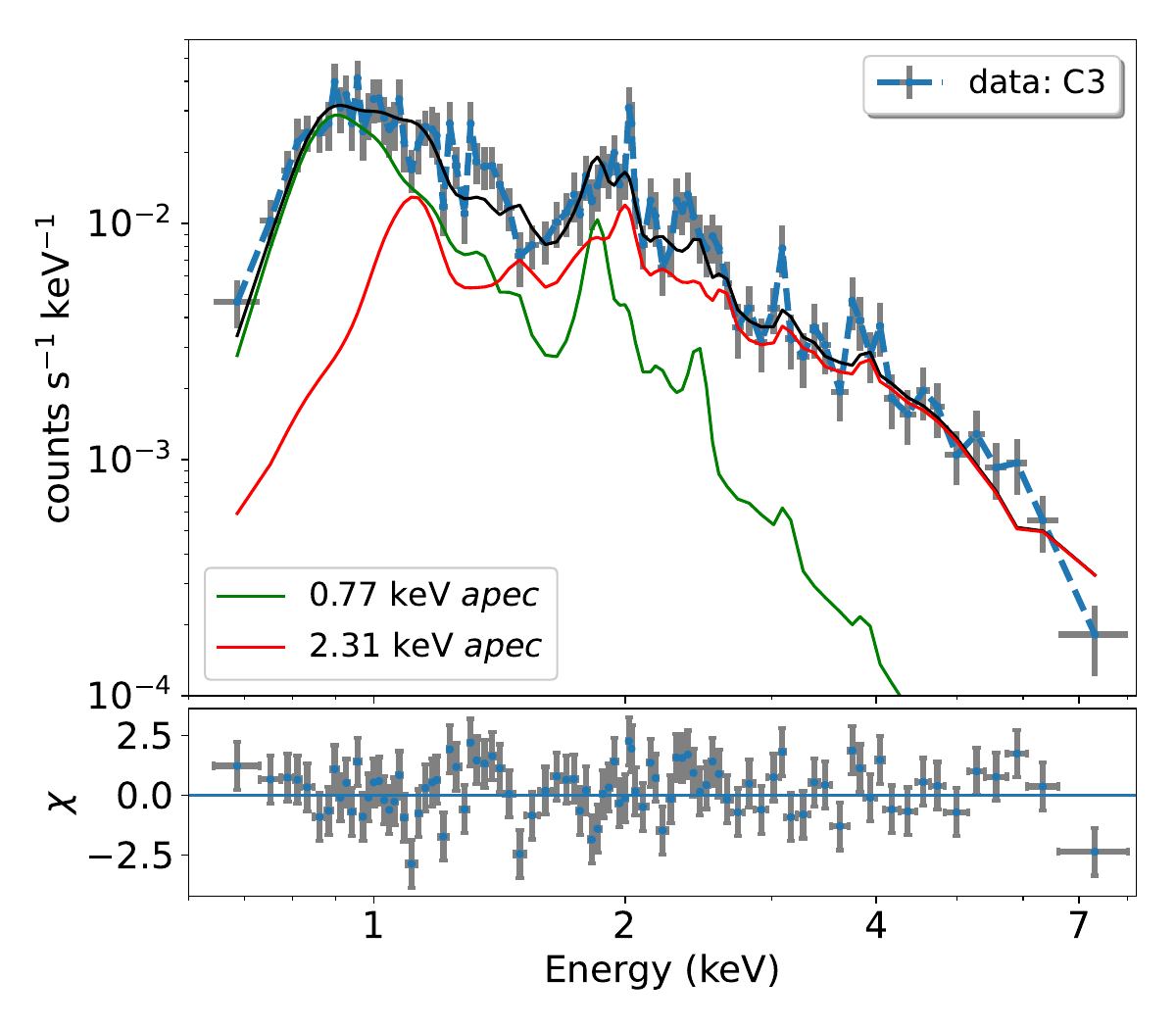}{0.33\textwidth}{(c) Data ID: C3, $\phi_{orb}= 0.16$} \label{fig:s3}}
\gridline{\fig{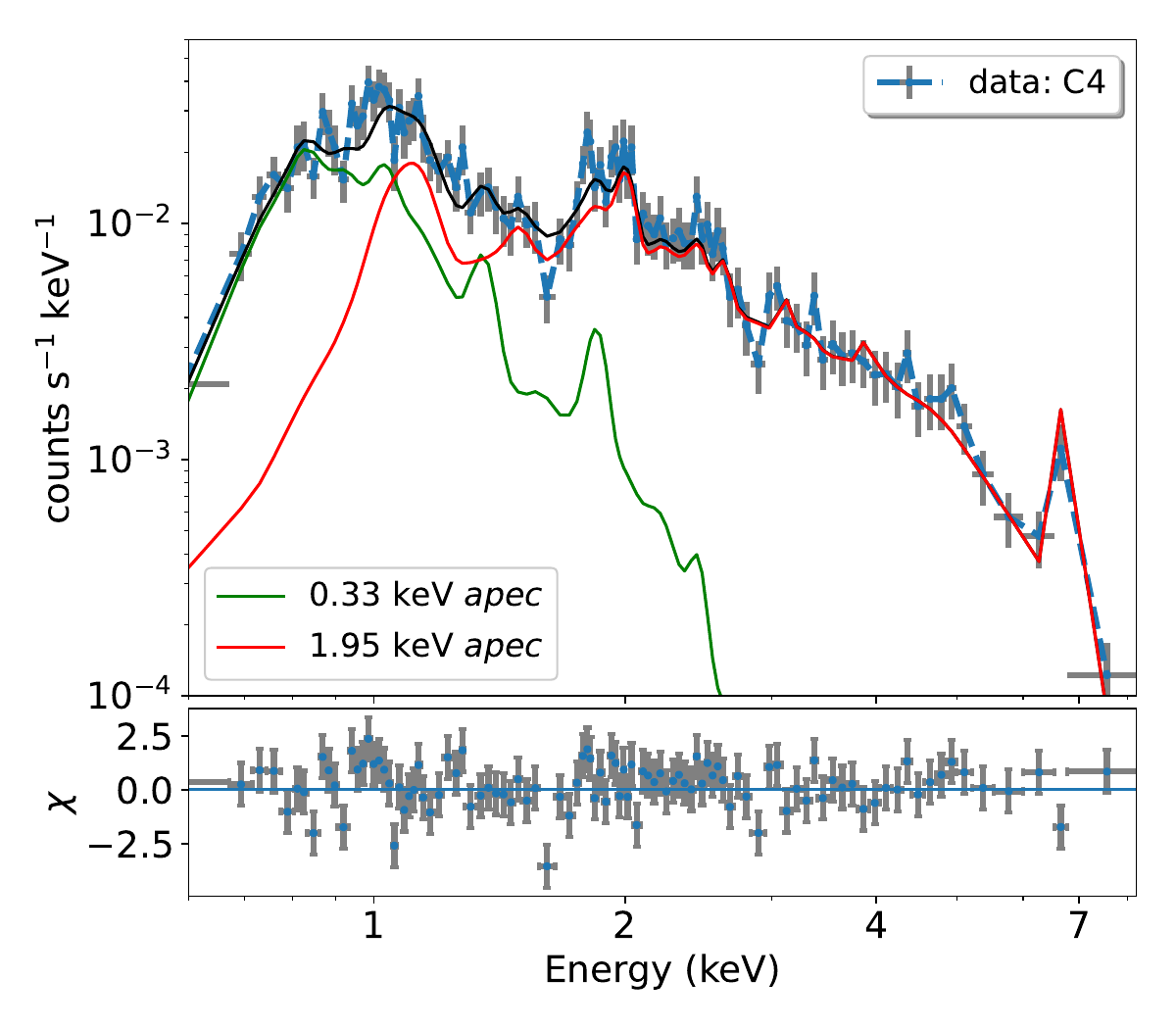}{0.33\textwidth}{(d) Data ID: C4, $\phi_{orb}= 0.29$} \label{fig:s4}
          \fig{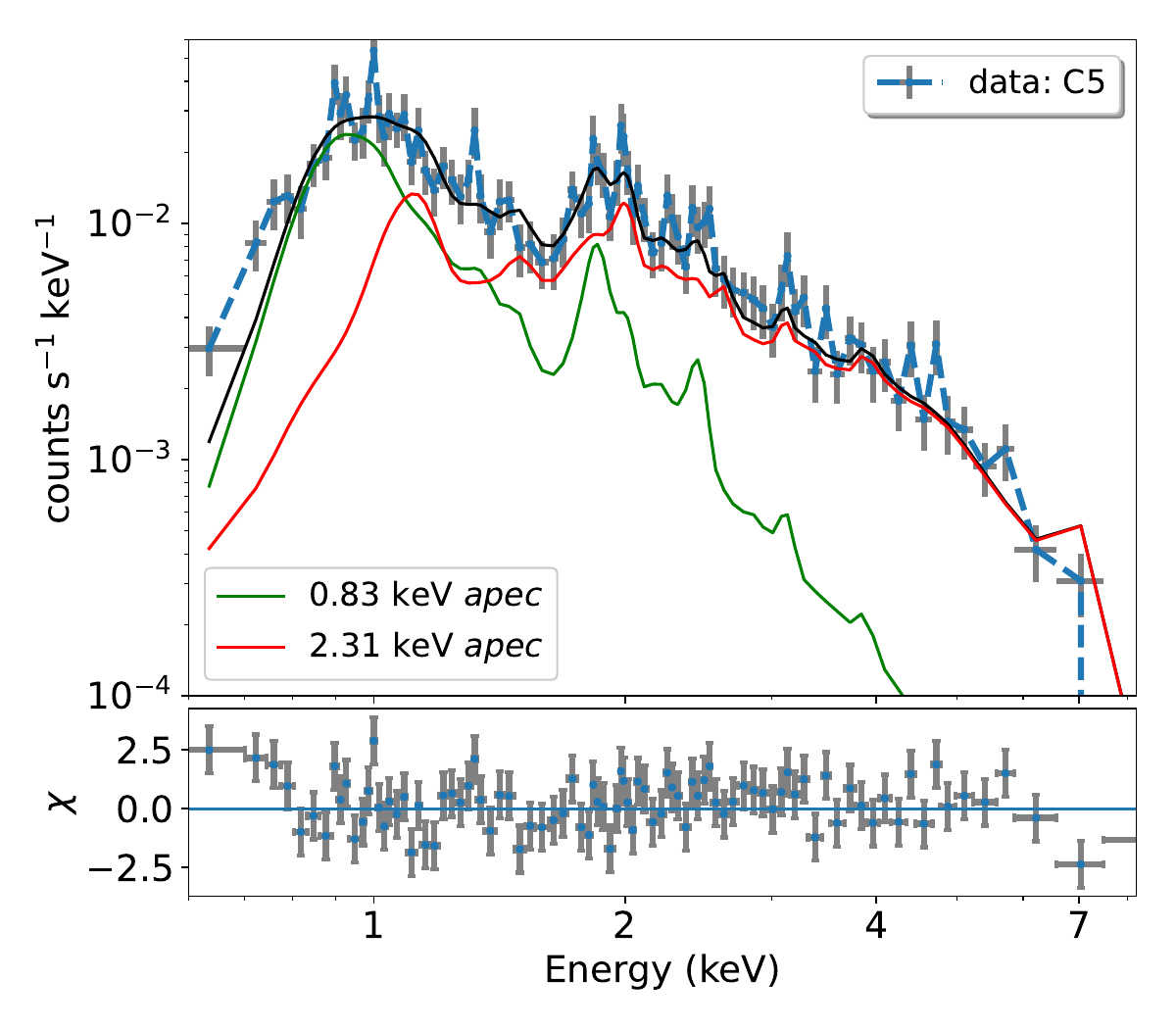}{0.33\textwidth}{(e) Data ID: C5, $\phi_{orb}= 0.38$} \label{fig:s5}
          \fig{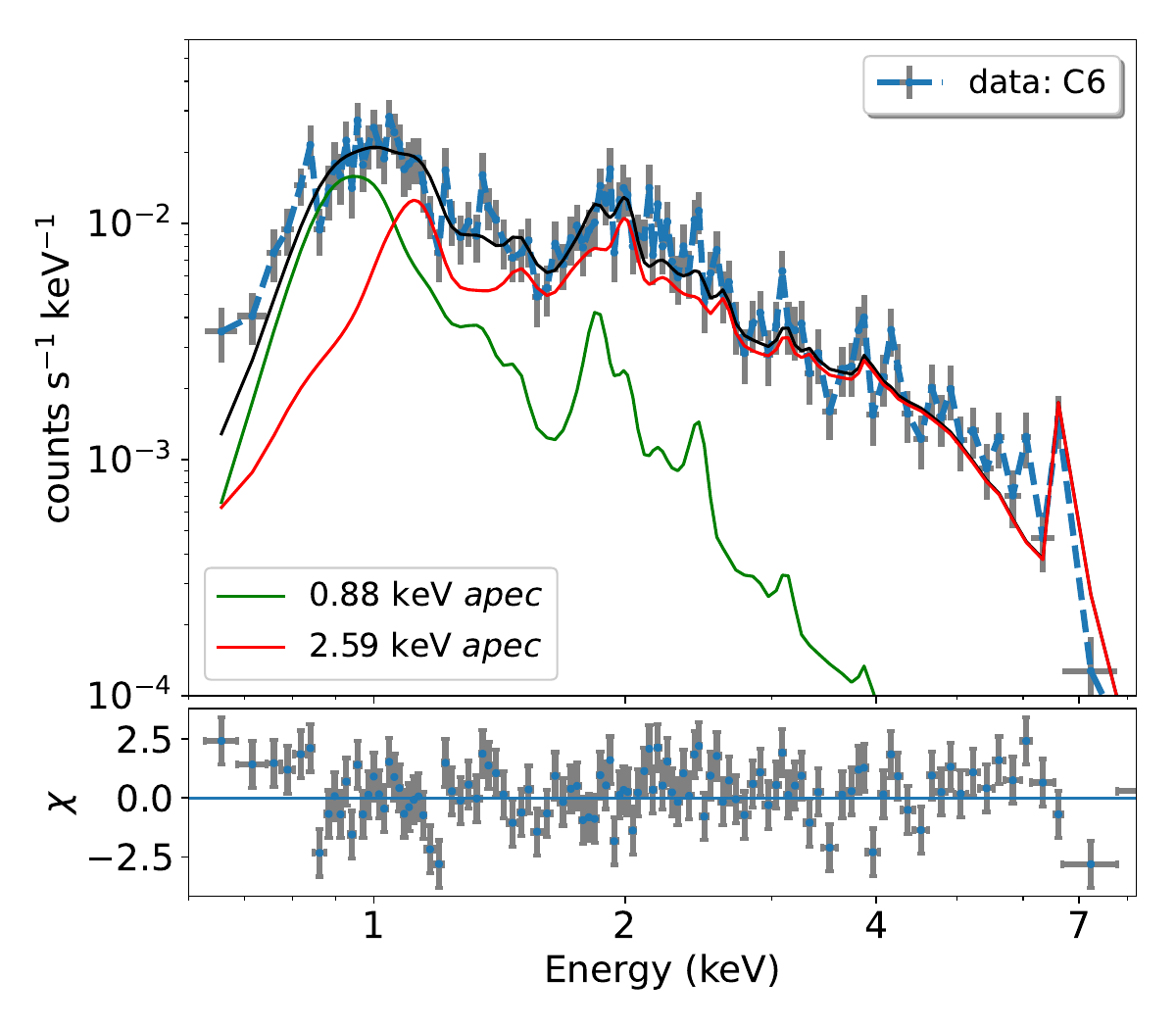}{0.33\textwidth}{(f) Data ID: C6, $\phi_{orb}= 0.50$} \label{fig:s6}}
\caption{Same as Fig. \ref{fig:example_spectra}, but for all the \cxo observations. \textbf{(a-f)} The panels show the \cxo ACIS spectra (in blue) along with the total model fit (black curve). In all cases the lower temperature model is shown in green, while the higher temperature is shown in red. For all figures, the bottom panel shows the value of $\chi$, defined as in Fig. \ref{fig:example_spectra}. \label{fig:all_spectra}}
\end{figure*}

In this section, we provide additional results of the spectral fitting of the \cxo and \xmm data. In Fig. \ref{fig:all_spectra}(a-f), we show the model fit to all six ACIS spectra. The best fit results from the spectral fitting using both 3T and 2T+pow models for all data are given in Table \ref{tab:spectra_fit_oths}.

\section{MCMC Analysis} \label{App:mcmc}

\begin{figure*}
\centering
\gridline{\fig{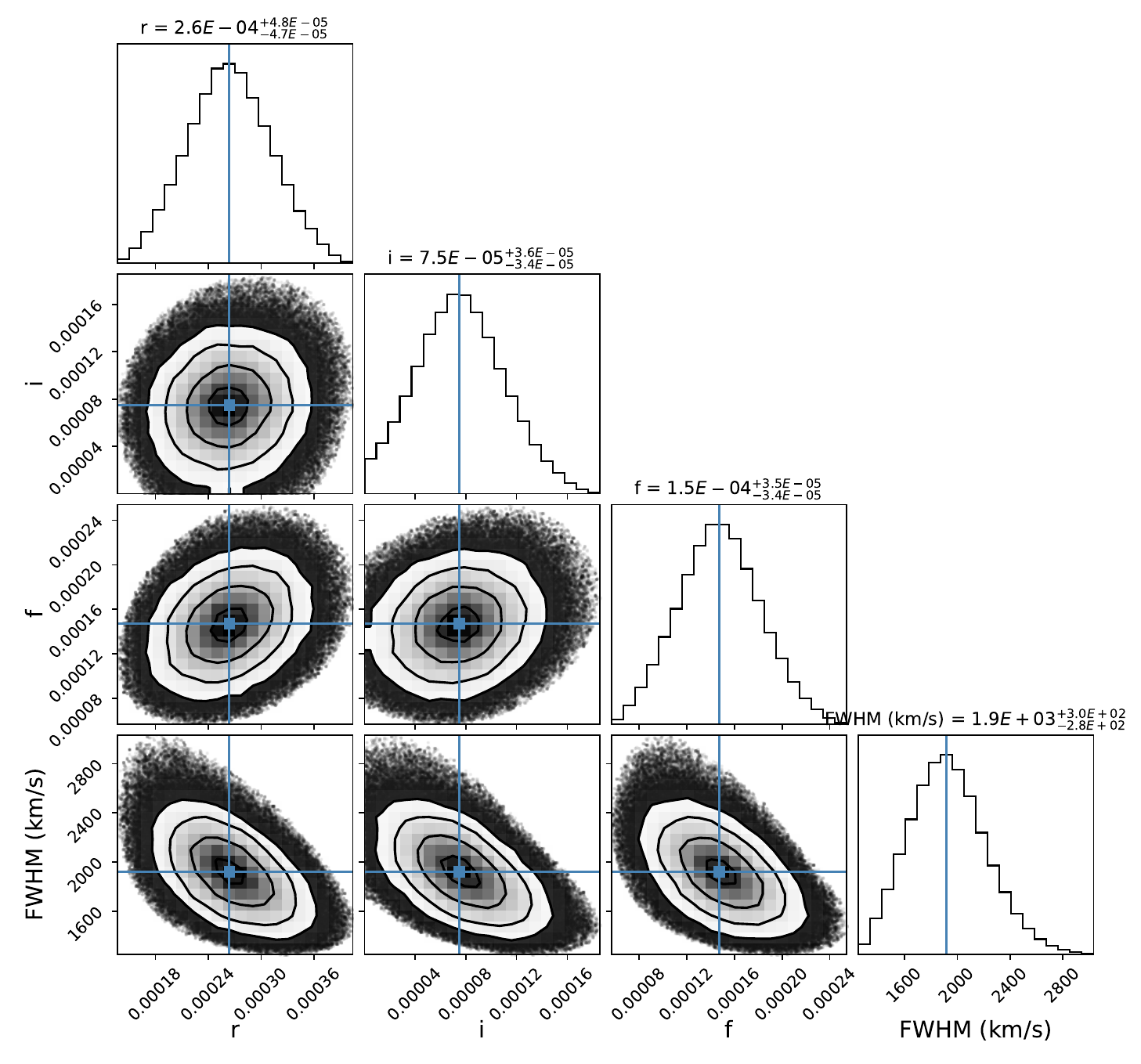}{0.465\textwidth}{(a) \ion{Si}{13}} \label{fig:mcmc_Plaskett_Si}
          \fig{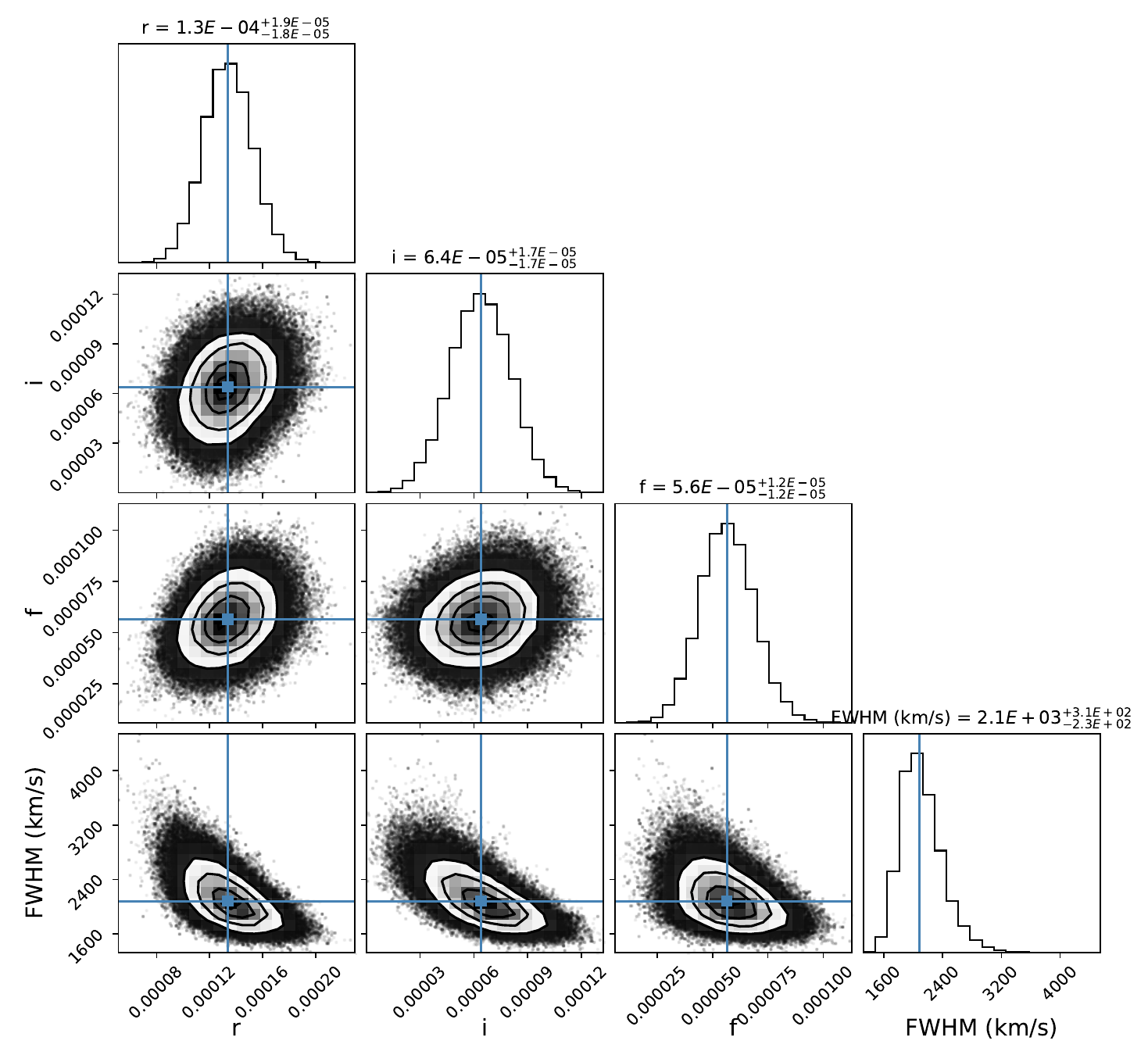}{0.465\textwidth}{(b) \ion{Mg}{11}} \label{fig:mcmc_Plaskett_Mg}}
    \caption{Corner plots corresponding to the 3-component Gaussian fit to the He-like triplets (left: \ion{Si}{13}; right: \ion{Mg}{11}). The labels $r, \ i, \ f$ correspond to different line components (recombination, $r$, inter-combination, $i$, and forbidden, $f$) and FWHM denotes the full-width half maxima of the line profiles. The median values are shown with blue lines.  }
    \label{fig:mcmc_fit}
\end{figure*}

In this section, we show the corner plots from the MCMC analysis while determining the $f/i$ ratios. The analysis was performed using the \textsc{Python}-based \textsc{emcee} package. For each analysis, we used a chain length of 20,000 and a burn length of 5,000. The corner plots for the \ion{Si}{13} and \ion{Mg}{11} lines are shown in Fig. \ref{fig:mcmc_fit}.

\begin{deluxetable*}{cccccccccc}
\tablecaption{Spectral fitting results obtained with \textit{xspec} from \cxo data similar to Table \ref{tab:2T_fit}, but with `3T', and `2T+pow' models. The $n_{\rm H}$ reported in the second column correspond to the additional absorption model, rather than the fixed ISM absorption. The numbers in parenthesis in the $\chi^2_{\rm reduced}$ column represents the number of degrees of freedom. T$_{\rm apec}$ and $\alpha$ represents the best-fit temperatures from \textit{apec} models, and the photon index obtained from power-law model, respectively. The normalization parameters are in units of $10^{-5}$ cm$^{-5}$.  All spectra were fitted in the energy range 0.2--10.0 keV. The flux values represented here are the unabsorbed flux after ISM correction.    \label{tab:spectra_fit_oths}} 
\tablehead{
\colhead{Data} &  \colhead{$n_{\rm H}$} & \colhead{$\chi^2_{\rm red}$}  &  \multicolumn{2}{c}{T$_{\rm apec}$ (keV)} & \colhead{$\alpha$} & \multicolumn{3}{c}{Norm, N ($\times 10^{-5}$ cm$^{-5}$)} & \colhead{$\log F_{(\rm 0.2-10 \ keV)}$}  \\
\colhead{ID}  &  \colhead{$\times 10^{22}$ cm$^{-2}$} & \colhead{} & \colhead{kT$_1$} & \colhead{kT$_2$} & \colhead{(or, kT$_3$)} & \colhead{$N_{\rm kT1}$} & \colhead{$N_{\rm kT1}$} & \colhead{$N_{\rm pow}$/$N_{\rm kT3}$} & \colhead{(erg cm$^{-2}$ s$^{-1}$)}
}
\startdata
\multicolumn{10}{c}{3T model} \\
C1 & 0.40 $\pm$ 0.13 & 1.14(110) & 0.33 $\pm$ 0.08 & 0.87 $\pm$ 0.07 & 2.5 $\pm$ 0.2 & 2.8 $\pm$ 2.2 & 0.7 $\pm$ 0.2 & 1.3 $\pm$ 0.1 & -11.497 $\pm$ 0.012 \\
C2 & 0.38 $\pm$ 0.13 & 1.22(82) & 0.24 $\pm$ 0.04 & 1.00 $\pm$ 0.08 & 2.4 $\pm$ 0.2 & 4.0 $\pm$ 3.9 & 0.9 $\pm$ 0.3 & 1.2 $\pm$ 0.2 & -11.497 $\pm$ 0.014 \\
C3 & 0.60 $\pm$ 0.12 & 1.14(85) & 0.22 $\pm$ 0.04 & 0.95 $\pm$ 0.11 & 2.2 $\pm$ 0.2 & 18.5 $\pm$ 15.7 & 0.9 $\pm$ 0.3 & 1.4 $\pm$ 0.2 & -11.489 $\pm$ 0.014 \\
C4 & 0.25 $\pm$ 0.13 & 1.07(95) & 0.35 $\pm$ 0.07 & 0.98 $\pm$ 0.17 & 2.1 $\pm$ 0.1 & 1.5 $\pm$ 1.5 & 0.3 $\pm$ 0.1 & 1.5 $\pm$ 0.1 & -11.503 $\pm$ 0.012 \\
C5 & 0.22 $\pm$ 0.08 & 1.31(83) & 56.76 $\pm$ 763.48 & 0.83 $\pm$ 0.04 & 2.2 $\pm$ 0.6 & 0.1 $\pm$ 0.2 & 0.9 $\pm$ 0.3 & 1.2 $\pm$ 0.4 & -11.548 $\pm$ 0.013 \\
C6 & 0.39 $\pm$ 0.15 & 1.21(105) & 0.21 $\pm$ 0.03 & 1.14 $\pm$ 0.15 & 2.7 $\pm$ 0.4 & 6.7 $\pm$ 7.1 & 0.6 $\pm$ 0.3 & 0.9 $\pm$ 0.2 & -11.566 $\pm$ 0.015 \\
\multicolumn{10}{c}{2T+pow model} \\
C1 & 0.08 $\pm$ 0.07 & 1.22(110) & 0.79 $\pm$ 0.04 & 2.48 $\pm$ 0.31 & 2.9 $\pm$ 0.5 & 0.5 $\pm$ 0.1 & 0.9 $\pm$ 0.2 & 0.4 $\pm$ 0.2 & -11.483 $\pm$ 0.013  \\
C2 &  0.34 $\pm$ 0.15 & 1.19(82) & 0.24 $\pm$ 0.04 & 1.07 $\pm$ 0.06 & 2.5 $\pm$ 0.2 & 3.2 $\pm$ 3.8 & 1.0 $\pm$ 0.3 & 0.6 $\pm$ 0.2 & -11.488 $\pm$ 0.014 \\
C3 & 0.60 $\pm$ 0.13 & 1.15(85) & 0.21 $\pm$ 0.04 & 1.01 $\pm$ 0.09 & 2.6 $\pm$ 0.2 & 19.4 $\pm$ 18.0 & 1.2 $\pm$ 0.3 & 0.7 $\pm$ 0.2 & -11.480 $\pm$ 0.014  \\
C4 & 0.31 $\pm$ 0.11 & 1.09(95) & 0.32 $\pm$ 0.04 & 1.73 $\pm$ 0.12 & 2.1 $\pm$ 0.6 & 2.7 $\pm$ 1.7 & 1.5 $\pm$ 0.3 & 0.1 $\pm$ 0.2 & -11.491 $\pm$ 0.012  \\
C5 & 0.04 $\pm$ 0.07 & 1.19(83) & 2.17 $\pm$ 0.52 & 0.88 $\pm$ 0.07 & 2.7 $\pm$ 0.3 & 0.5 $\pm$ 0.2 & 0.4 $\pm$ 0.1 & 0.5 $\pm$ 0.2 & -11.489 $\pm$ 0.016  \\
C6 & 0.39 $\pm$ 0.12 & 1.18(105) & 0.20 $\pm$ 0.02 & 1.28 $\pm$ 0.08 & 2.2 $\pm$ 0.2 & 7.4 $\pm$ 6.4 & 0.9 $\pm$ 0.2 & 0.3 $\pm$ 0.1 & -11.554 $\pm$ 0.015  \\
\enddata
\end{deluxetable*}

\end{document}